\documentclass{svjour3}

\usepackage{macros}
\usepackage{lstcoq}
\usepackage{lstocaml}

\usepackage{amsfonts,stmaryrd,amsmath}

\usepackage{graphics}
\usepackage{array}
\usepackage{paralist}
\usepackage[all]{xy}

\usepackage{caption}
\usepackage{subcaption}
\usepackage{multicol}

\usepackage{tikz}
\usetikzlibrary{trees}
\usepackage{microtype}

\usepackage{hyperref}
\hypersetup{pdftitle={Implementing and reasoning about hash-consed data structures in Coq},
  pdfauthor={Thomas Braibant, Jacques-Henri Jourdan and David Monniaux}}

\journalname{Journal of Automated Reasoning}

\author{Thomas~Braibant \and Jacques-Henri~Jourdan\and David~Monniaux}
\institute{
  T.~Braibant \and J.-H.~Jourdan \at
  \href{http://www.inria.fr/}{Inria} Paris-Rocquencourt,
  Domaine de Voluceau, BP 105, 78153 Le Chesnay, France \\
  \email{thomas.braibant@inria.fr, jacques-henri.jourdan@inria.fr}
\and
  D.~Monniaux \at
  \href{http://www.cnrs.fr/}{CNRS} / \href{http://www-verimag.imag.fr/}{VERIMAG},
  Centre \'Equation,
  2, avenue de Vignate,
  38610 Gi\`eres, France \\
  \email{david.monniaux@imag.fr}
}

\title{Implementing and reasoning about hash-consed data structures in Coq
\thanks{The research leading to these results has received funding from the \emph{Agence nationale de la recherche} (ANR) under project
  ``\href{http://verasco.imag.fr/}{VERASCO}'' and the
 \href{http://erc.europa.eu/}{European Research Council}
 under the European Union's Seventh Framework Programme (FP/2007-2013) / ERC Grant Agreement nr. 306595
  ``\href{http://stator.imag.fr/}{STATOR}''}}

\usepackage[backend=bibtex,style=numeric-comp]{biblatex}
\bibliography{bib}
\begin{document}

\date{November 2013}

\maketitle

\begin{abstract}
  We report on four different approaches to implementing hash-consing
  in Coq programs. The use cases include execution inside Coq, or
  execution of the extracted OCaml code. We explore the different
  trade-offs between faithful use of pristine extracted code, and code
  that is fine-tuned to make use of OCaml programming constructs not
  available in Coq.
  We discuss the possible consequences in terms of performances and
  guarantees.
  We use the running example of binary decision diagrams and then
  demonstrate the generality of our solutions by applying them to
  other examples of hash-consed data structures.
\end{abstract}

\section{Introduction}
\emph{Hash-consing} is a programming technique used to share identical
immutable values in memory, keeping a single copy of semantically
equivalent objects.  It is especially useful in order to get a compact
representation of abstract syntax trees.  A hash-consing library
maintains a global pool of expressions and never recreates an
expression equal to one already in memory, instead reusing the one
already present.  In typical implementations, this pool is a global
hash-table.  Hence, the phrase \emph{hash-consing} denotes the
technique of replacing node \emph{cons}truction by a lookup in a hash
table returning a preexisting object, or creation of the object
followed by insertion into the table if previously nonexistent.  This
makes it possible to get \emph{maximal sharing} between objects, if
hash-consing is used systematically when creating objects.

Moreover, a unique identifier is given to each object, allowing fast
hashing and comparisons.
This makes it possible to do efficient \emph{memoization} (also known
as \emph{dynamic programming}): the results of an operation are
tabulated so as to be returned immediately when an identical
sub-problem is encountered.

Hash-consing and memoization are examples of imperative
techniques that are of prime importance for performance.
The typical way of implementing hash-consing (a global hash table)
does not translate easily into Coq. The reason is that the Gallina
programming language at the heart of the Coq proof assistant is a
purely applicative language, without imperative traits such as hash
tables, pointers, or pointer equality.
In the following, we discuss how hash-consing can be implemented using
the Coq proof assistant (without modifying it) with two possible use
cases in mind:
\begin{itemize}
\item execution inside {Coq} with reasonable efficiency, e.g. for
  proofs by reflection;
\item efficient execution when extracted to {OCaml},
  e.g. for use in a model-checking or static analysis tool proved
  correct in Coq.
\end{itemize}

More precisely, we present a few \emph{design patterns} to implement
hash-consing and memoization on various examples, and we use binary
decision diagrams (BDDs) as our prime example.
We evaluate these design patterns based on several criterion. For
instance, we discuss for each pattern how we handle the maximal
sharing property and how easy it is to prove the soundness of our
implementation. We also discuss whether or not it is possible to
compute inside Coq with a given solution.

\paragraph{Reading this article.} In the following, we will try to
separate some implementation details or Coq specific comments from the
main text of the article.
These remarks, titled ``For the practical mind'', can be safely skipped
on the first reading.

\paragraph{Outline.} This paper is organized as follows.
Since hash-consing can be seen as a particular kind of memoization,
where one memoizes constructor applications, we begin this paper with
a short review of the techniques that are available in Coq in order to
memoize functions (\secref{sec:memo}).
Then, we give a short primer on BDDs in \secref{sec:bdd} and present
the ``menu'' of our four implementations of BDD libraries in
\secref{sec:menu}. These implementations are subsequently described in
\secref{sec:pure} and \secref{sec:impure} and use different design
patterns.
We discuss the pros and cons of each design pattern in
\secref{sec:discussion}.
We adapt these design patterns to implement strong reduction of
hash-consed lambda-terms in \secref{sec:lambda}.
We discuss related work and conclude this paper in \secref{sec:conclusion}.

\section{Memoization in Coq: state of the art}\label{sec:memo}
Generally speaking, there are two ways to use Coq code: either to
execute it inside Coq, e.g., to define reflective decision procedures;
or to extract it to OCaml (or other supported languages), then compile the
extracted code.
These two usages impose constraints on the possible design
patterns. In the remainder of this section, we describe what the state
of the art in memoization approaches is with respect to these two use
cases.

\subsection{State monad}
\label{ssec:state_monad}
The first idea to implement memoization in Coq is to use a dedicated
data structure to tabulate function calls: that is, use a finite map
of some sort to store the pairs \emph{(argument, result)} for a given
function.  Users of a library implemented in this fashion must, then,
thread this state through the program using a state monad.

While this solution always works, it has non-negligible verification
cost: we must prove that the bindings in the memoization table are
correct, i.e., the values held in the table correspond to the result
of the would-be corresponding function calls.

Then, if we memoize a function of type \coqe{A -> B}, we get a
function of type \coqe{A -> M B}, where \coqe{M} is the type
constructor associated with our state monad. Even if the latter seems
equivalent to the former, these types are different: therefore, this
tiny modification (memoizing a function) requires modifying every
caller of the memoized function to use a monadic style.

Since it is cumbersome to perform all computations inside a state
monad, we continue this section with a review of other (partial)
solutions to the memoization problem.

\subsection{Shallow memoization}
% marche pour fonctions dont le type n'est pas arbitraire.
% marche pour accesseurs assez simples.

\begin{figure}
\begin{multicols}{2}
\begin{coq}
Section t.
  Context {A : Type}.
  CoInductive lazy : Type :=
  | Thunk : A -> lazy.

  CoInductive lazies : Type :=
  | Cons : lazy -> lazies -> lazies.

  Fixpoint nth n cst :=
    match n, cst with
      | O, Cons (Thunk xi) _ => xi
      | S p, Cons _ c => nth p c
    end.

  CoFixpoint mk_lazy {B} f (n : B) :=
    Thunk (f n).

  CoFixpoint memo' f n :=
    Cons (mk_lazy f n) (memo' f (S n)).

  Definition memo f := memo' f 0.
End t.

(* A costly version of the identity *)
Fixpoint big {A} n : A -> A:=
match n with
  | 0 => fun x => x
  | S n =>fun x =>  big n (big n x)
end.

Definition id n := big n n.
Definition k := memo id.
Definition run x := nth x k.
\end{coq}
\end{multicols}
  \caption{Memoization using co-inductives}
  \label{fig:memo-coinductive}
\end{figure}

It is possible in some cases to make a \emph{shallow embedding} of
this memoization table in Coq.
\Textcite{Mel12} describes how to use co-inductive objects as a
cache that stores previously computed values of a given function. In
this section, we present his clever idea in
\figref{fig:memo-coinductive}, with a few cosmetic changes\footnote{We
  found out that similar definitions actually made their way into
  Coq's standard library under the name {\tt StreamMemo.v}, yet we
  believe that this idea deserves more explanation than what is
  available.}.
A \coqe{lazy} value is defined as a co-inductive, which
will effectively be represented as a thunk when executed by Coq's
virtual machine, and will be extracted to OCaml as a \ocamle{Lazy.t} thunk.
A \emph{thunk} is a term whose evaluation is frozen until it is actually
needed (lazy evaluation), then the computed value is cached in the thunk
so that it is instantly returned in case the term is evaluated again
(thus, performing a kind of memoization).

Then, \coqe{lazies} represents a stream (an infinite list) of thunks,
that can be peeked at using \coqe{nth}. Each
time the virtual machine has to destruct the \coqe{cst} argument of
\coqe{nth}, it will remember the two arguments of this cons
cell. Then, it suffices to build a stream \coqe{memo} that computes
the value of a function \coqe{f} for each possible natural number: in
practice, \coqe{f} will be evaluated lazily, and only once per input
value.

As an example, the following snippet of code memoizes the
time-consuming function \coqe{id}.
(Remark that it also demonstrates that the intermediate calls of
\coqe{id} are not memoized.)
\begin{multicols}{2}\begin{coq}
Time Eval vm_compute in run 26. (* 6s *)
Time Eval vm_compute in run 26. (* 0s *)
Time Eval vm_compute in run 25. (* 3s *)
Time Eval vm_compute in run 25. (* 0s *)
Time Eval vm_compute in id 25.  (* 3s *)
Time Eval vm_compute in id 25.  (* 3s *)

$ $
\end{coq}\end{multicols}

Note that this clever trick is not limited to functions over
Peano numbers, and could be adapted for branching structures,
e.g. memoizing functions over binary numbers.
Unfortunately, it seems hard to adapt this idea to memoize recursive
calls: consider for instance the function
\begin{coq}
Fixpoint exp n := match n with 0 => 1 | S n => exp n + exp n end.
\end{coq}
In order to compute \coqe{exp} using a linear number of recursive
calls we need a memoizing fixed-point combinator rather than a way to
memoize a predefined function.

% NOTE DM: pas super clair comme distinction
(To be complete, let us add that functions \coqe{f : nat -> A} that can
be expressed as the iteration of a given function \coqe{g : A -> A}, can
actually be memoized using the aforementioned technique. While we
obviously could rewrite \coqe{exp} to fit into that scheme, this is
not the case of the other functions that we use through our developments.)

This problem is somewhat representative of the one that we have to
face when it comes to BDDs: first, memoizing fixed-point combinators
are crucial in our work; second, the function that we memoize depends
on the global state of the hash-consing machinery, which evolves
through times. In both cases, we fall out of the scope of the above
trick: there is no predefined function to memoize.

\subsection{Memoization in extracted code}
\label{ssec:adjustable-references}
Now, we turn to a partial solution to the memoization problem, that
works when the user is solely interested in executing the code
extracted from a Coq program.

\Textcite{adjustable-references} introduced recently a restricted form of mutable state called \emph{adjustable references}.
The idea is that adjustable references store some internal value that
can only be updated in innocuous ways. That is, adjustable references
are equipped with an observation function; and the update function
ensures that the result of the observations remain equals through
updates.
It is therefore possible to adjust the values held in the references
in a way that changes, e.g., the costs of some computations, yet does
not change the result of these computations.

Vafeiadis demonstrated how to use adjustable references to memoize a
function \coqe{f} by tabulating its results: the adjustable reference
internal state is a finite map, while the observation function simply
returns \coqe{f}. Updating the contents of the finite map does not
change the observation function; but subsequent reads may make use of
the fact that a value was previously computed and stored inside the
memoization cache.
We refer the reader to Vafeiadis' article for more details about this idea.

Remark that it is again the case that this technique does not scale to
the definition of memoizing fixed-point combinators. Therefore,
adjustable references do not quite fit our needs either.

\subsection{Summary}
Out of the three memoization techniques available in Coq that we
described, only the first one suits our need: it makes it possible to
define memoizing fixed-point combinators. In this paper, we use
this \emph{state monad} approach, then we will come up with a new idiom.

\section{A primer on binary decision diagrams}\label{sec:bdd}
In the following, we use ``reduced ordered binary decision diagrams''
(ROBDDs, BDDs for short) as a running example of hash-consed
data structures.
BDDs are representations of Boolean functions and are
often used in software and hardware formal verification tools, in
particular \emph{model checkers}~\cite{taocp-bdd}.

\paragraph{The data structure.} A Boolean function $f: \{0,1\}^n \rightarrow \{0,1\}$ can be
represented as a complete binary tree with $2^n-1$ decision nodes,
labeled by variables $x_i$ according to the depth $i$ from the root
(thus the adjective \emph{ordered}); edges are labeled $0$ and $1$;
leaves are labeled \coqe{T} (for true) or \coqe{F} (for false).  The
semantics of such a diagram is: $f(x_1,\dots,x_n)$ is obtained by
traversing from the root and following the $0$ or $1$ edge of a node
labeled by $x_i$ according to the value of $x_i$.

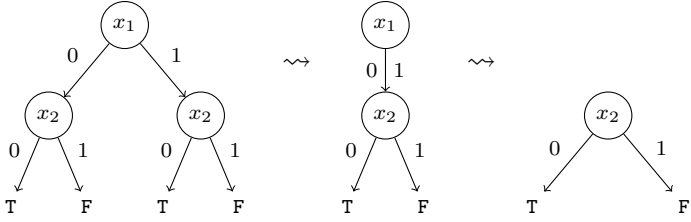
\begin{figure}
\begin{center}
\begin{tikzpicture}[level/.style={level distance=12mm,sibling distance=20mm/#1}]
\node[circle,draw] (t) {$x_1$}
  child {
    node[circle,draw] (t0) {$x_2$}
      child { node (t00) {\texttt{T}} edge from parent[->] node[above left] { $0$ } }
      child { node (t01) {\texttt{F}} edge from parent[->] node[above right] { $1$ } }
    edge from parent[->] node[above left] { $0$ }
  }
  child {
    node[circle,draw] (t1) {$x_2$}
      child { node (t10) {\texttt{T}} edge from parent[->] node[above left] { $0$ } }
      child { node (t11) {\texttt{F}} edge from parent[->] node[above right] { $1$ } }
    edge from parent[->] node[above right] { $1$ }
  }
  ;
\end{tikzpicture}
\quad\raisebox{2cm}{\large$\leadsto$}\quad
\begin{tikzpicture}[level/.style={level distance=12mm,sibling distance=20mm/#1}]
\node[circle,draw] (t) {$x_1$}
  child {
    node[circle,draw] (t0) {$x_2$}
      child { node (t00) {\texttt{T}} edge from parent[->] node[above left] { $0$ } }
      child { node (t01) {\texttt{F}} edge from parent[->] node[above right] { $1$ } }
    edge from parent[->] node[left] { $0$ } node[right] { $1$ }
  }
  ;
\end{tikzpicture}
\quad\raisebox{2cm}{\large$\leadsto$}\quad
\begin{tikzpicture}[level/.style={level distance=12mm,sibling distance=20mm/#1}]
\node[circle,draw] (t0) {$x_2$}
      child { node (t00) {\texttt{T}} edge from parent[->] node[above left] { $0$ } }
      child { node (t01) {\texttt{F}} edge from parent[->] node[above right] { $1$ } };
\end{tikzpicture}
\end{center}
\caption{The BDD for the function $f(0,0)=\texttt{T}$, $f(0,1)=\texttt{F}$,
$f(1,0)=\texttt{T}$, $f(1,1)=\texttt{F}$}\label{fig:bdd-example}
\end{figure}

Such a tree can be \emph{reduced} by merging identical subtrees, thus
becoming a connected \emph{directed acyclic graph} (DAG; see second
diagram in \figref{fig:bdd-example}); furthermore, decision nodes with
identical children are removed (see third diagram in
\figref{fig:bdd-example}). These transformations preserve
semantics. The reduced representation is \emph{canonical}: given a
variable ordering $x_1,\dots,x_n$, a function is represented by a
unique ROBDD.

\paragraph{Building BDDs.} In practice, one directly constructs the
reduced tree; let us see how.  All BDD nodes allocated so far are
stored in a global hash table, so that a new object is created only if
not already in the hash table: \emph{there are never two identical
  objects at two different locations in memory}, a crucial invariant
that must be maintained throughout the execution.

Each BDD node is given a unique identifier: for instance, the current
value of a global 64-bit counter incremented following each
allocation\footnote{In certain implementations, the unique identifier is the
  address of the node; this supposes that objects never change
  address, which is not the case in OCaml.}.  Since a new object is
never created if an identical object already exists, two objects are
equal if and only if they have the same unique identifier.  The
equality test, which is usually expensive over tree structures, since
it requires full traversal, is instead implemented by a very fast
comparison of unique identifiers.

Unique identifiers are also used for hashing. When attempting to
construct a node $(v,b_0,b_1)$ where $v$ is the variable, $b_0$ is the
subtree labeled $0$ and $b_1$ is the subtree labeled $1$, one computes
the hash value of the node as $H(v,u_0,u_1)$ where $u_0$ and $u_1$ are
the respective unique identifiers of the subtrees $b_0$ and $b_1$.
This hash value is then used to look up the node in the hash table.
Again, such shallow hashing is considerably faster than having to
traverse the tree structures.

\paragraph{Operations on BDDs.} The principal way to build BDDs is to
combine the diagrams of two functions $f$ and $g$ in order to obtain
the BDD for other functions such as $f \wedge g$, $f \vee g$ of $f
\oplus g$. The basic combinator of BDDs is called node
\emph{melding}~\cite{taocp-bdd}.
Intuitively, melding corresponds to a traversal of the internal nodes
of two BDDs to build a third diagram. This traversal respects the
order on variables, and thus, the third diagram will naturally be
ordered. Then, all binary operations on BDDs are defined as one
particular instance of melding.

\newcommand \meld {\diamond}

Suppose that we have two nodes $\alpha = (v,l,h)$ and $\alpha' =
(v',l',h')$. The melding of $\alpha$ and $\alpha'$, denoted by $\alpha
\meld \alpha'$ is defined as follows:
\begin{displaymath}
  \alpha \meld \alpha' =
  \begin{cases}
    (v, l \meld l', h \meld h') & \mbox{if } v = v' \\
    (v, l \meld \alpha', h \meld \alpha') & \mbox{if } v < v' \\
    (v', \alpha \meld l', \alpha \meld h') & \mbox{if } v > v'
  \end{cases}
\end{displaymath}
Then, the binary operation $\meld$ is entirely defined by the results
of
\begin{center}
  \renewcommand\top{{\tt T}}
  \renewcommand\bot{{\tt F}}
  \begin{tabular}{cccc}
    $\bot \meld \alpha$ &
    $\alpha \meld \bot$ &
    $\top \meld \alpha$ &
    $\alpha \meld \top$
  \end{tabular}
\end{center}
For instance, the conjunction operation $f \wedge g$ can be defined by
melding the BDDs for $f$ and $g$, and using the following rewrite
rules for the leaf cases:
\begin{center}
  \renewcommand\top{{\tt T}}
  \renewcommand\bot{{\tt F}}
  \begin{tabular}{cccc}
    $\bot \meld \alpha \to \bot$ &
    $\alpha \meld \bot \to \bot$ &
    $\top \meld \alpha \to \alpha$ &
    $\alpha \meld \top \to \alpha$
  \end{tabular}
\end{center}
For the sake of clarity, we focus on binary operations in the
following: they are representative of the difficulties we had to
face\footnote{Note that we only need two rewrite rules for commutative
  binary operations.}.

\paragraph{Memoization.}
The fact that each node has a unique identifier also makes it possible
to memoize the results of BDD operations. One keeps a map from
sub-problems (a pair of nodes $\alpha$ and $\alpha'$) to nodes (the
result of $\alpha \meld \alpha'$), so as to be returned immediately
when a previously solved sub-problem is encountered.

In a BDD library, memoization is crucial to implement the or/and/xor
operations with time complexity in $O(|a|.|b|)$ where $|a|$ and $|b|$
are the sizes of the inputs: as the function is executed, its results
on the subtrees of the original problem are stored in a structure
indexed by $(u_a,u_b)$ where $u_a$ and $u_b$ are the unique
identifiers of the $a$ and $b$ inputs.  In contrast, the naive
approach has exponential complexity, since the function may be
evaluated exponentially often on the same couple of subtrees.

\section{Implementing BDDs in Coq.}\label{sec:menu}
The following two sections \secref{sec:pure} and \secref{sec:impure}
describe a total of \emph{four} implementations of BDDs in Coq. To
make things clear for the reader, we give each of these implementation
a reference name, a pointer to the relevant section of the paper and a
short description.
\begin{description}
\item[\sc pure-deep.] See \secref{sec:pure-deep}. A pure Coq
  implementation of BDDs that makes a deep embedding of memory as
  finite maps and uses indices as surrogates of pointers.
\item[\sc pure-shallow.] See \secref{sec:pure-shallow}. A pure Coq
  implementation of BDDs that uses a shallow embedding of memory.
\item[\sc smart.] See \secref{sec:smart}. An ''impure'' implementation
  of BDDs in Coq: we implement hash-consing and memoization through the
  extraction mechanism of Coq.
\item[\sc smart+uid.] See \secref{sec:smart+uid}. A variation on the
  previous approach in which we discuss how to expose and axiomatize
  the operations on the unique identifiers associated with BDD nodes.
\end{description}

\section{Pure solutions}\label{sec:pure}
As the reader can gather from the summary above, we describe two pure
Coq implementations of BDDs libraries in this section, that differ
mainly by the way we model the memory and the allocation of nodes.

\subsection{The {\sc pure-deep} approach}\label{sec:pure-deep}
The idea here is to model the memory using finite maps inside Coq and
use indices in the maps as surrogates for pointers, implementing BDD
operations as manipulations of these persistent maps. Such an
implementation\footnote{One should also mention the seminal work by
  E. Ledinot in 1993 on the canonicity of binary decision dags,
  available from the Coq contributions.} was described in
\cite{DBLP:conf/asian/VermaGPA00,verma:inria-00072797}, but we propose
a new one here for the sake of completeness, and because the old one
did not age well w.r.t. the evolution of Coq. Our implementation is
defined as follows.

First, we assign a unique identifier to each decision node.
Second, we represent the directed acyclic graph underlying a BDD as a
Coq finite map from identifiers to decision nodes (that is, tuples
that hold the left child, the node variable and the right child).
For instance, the following graph, on the left, can be represented using
the map on the right.

\begin{tabular}{cc}
  \begin{minipage}{0.5\linewidth}
    \begin{displaymath}
      \SelectTips{eu}{10}
      \xymatrix@R=0.5pc @C=0.5pc{
        x_1 \ar@{->}[dr] \ar@/_1pc/@{->}[dddr] & & & \\
        & x_2 \ar@{->}[dd] \ar@{->}[dr] & & \\
        & & x_3 \ar@{->}[dl] \ar@{->}[dr] & \\
        & F &  & T
      }
    \end{displaymath}
  \end{minipage}
  &
  \begin{tabular}{c@{$\quad\mapsto\quad$}l}
    1 & (F, $x_1$, N 2) \\
    2 & (F, $x_2$, N 3) \\
    3 & (F, $x_3$, T)
  \end{tabular}
\end{tabular}

\noindent Then, we implement the hash-consing pool using another map
from decision nodes to node identifiers and a \coqe{next} counter that
is used to assign a unique identifier to a fresh node.
In the situation above, \coqe{next} is equal to 4 and the hash-consing map
is defined as follows
\begin{center}
  \begin{tabular}{c@{$\quad\mapsto\quad$}l}
     (F, $x_1$, N 2)& 1 \\
     (F, $x_2$, N 3)& 2\\
     (F, $x_3$, T)  & 3
   \end{tabular}
\end{center}

\newcommand\fmap{\leadsto}

\paragraph{Notations.}
Through this paper, we stay as close as possible to our Coq code. Yet
we allow us some liberty when it comes to finite maps (that are
pervasive in our code): the notation \coqe{t $\fmap$ s} denotes a
type of efficient maps from type \coqe{t} to type \coqe{s}; and we use
indiscriminately the notations \coqe{get} and {\tt set} that
respectively access and update such finite maps. Implicitly, these two
notations have the following types:
\begin{center}
  \begin{tabular}{rcl}
    {\tt get} & : & {$\tt A \to (A \fmap B) \to option~B$ } \\
    {\tt set} & : & {$\tt A \to B \to (A \fmap B) \to (A \fmap B)$ }
  \end{tabular}
\end{center}
(Note that we do use the efficient finite maps from the Coq standard
library; but we chose here to abstract from the particular module
definitions and names that we have to use in our code, for the sake of
legibility.)

\newenvironment{practical}{\paragraph{For the practical mind.}}{\hfill$\Box$}

\begin{practical}
  In the following, we use Coq's positive numbers \coqe{positive} as
  unique identifiers, and use Patricia trees for finite maps.
\end{practical}

\medskip

\figref{fig:pure-hashcons} shows our inductive definitions and the
code of the associated allocation function \coqe{mk_node}, knowing
that \mbox{\coqe{alloc n st}} allocates the fresh node \coqe{n} in the
hash-consing state \coqe{st} (taking care of updating both finite maps
and incrementing the ``next fresh'' counter).  Then, equality between
BDDs (\coqe{eqb}) is provided by decidable equality over node
identifiers.

\begin{figure}[t]
  \centering

\begin{coq}
Inductive expr := F | T | N : positive -> expr.

Definition eqb a b :=
  match a,b with | T,T => true | F,F => true | N x, N y => (x =? y) | _, _ => false end.

Definition node := (expr * var * expr).

Record hashcons_state := {
  graph : positive $\fmap$ node;
  hmap  : node $\fmap$ positive;
  next  : positive
}.

Definition alloc u st :=
  (st.(next), {|
    graph := set st.(next) u st.(graph);
    hmap  := set u st.(next) st.(hmap);
    next  := st.(next) + 1
  |}).

Definition mk_node (l : expr) (v : var) (h : expr) st :=
  if eqb l h then (l,st)
  else match get (l,v,h) st.(hmap) with
         | Some x => (N x, st)
         | None => let (x, st) := alloc (l,v,h) st in (N x, st)
       end.
\end{coq}

  \caption{Hash-consing in pure Coq, using a deep-embedding}\label{fig:pure-hashcons}
\end{figure}

\begin{figure}[t]
\centering
\begin{coq}
Record incr (st1 st2 : hashcons_state) : Prop := {
  incr_lt : st1.(next) $\le$ st2.(next);
  incr_find : forall p x, get p st1.(graph) = Some x -> get p st2.(graph) = Some x
}.
\end{coq}
\caption{Monotonicity predicate over hash-consing states}\label{fig:pure-incr}
\end{figure}

All operations thread the current global state in a monadic
fashion.
The correctness of BDD operations corresponds to the facts that
\begin{inparaenum}[1)]
\item the global state is used in a monotonic fashion (that is the
structure of the resulting global state is a refinement of the input
one, see \figref{fig:pure-incr});
\item the resulting global state is well-formed;
\item the denotation of the resulting BDD expression is correct.
\end{inparaenum}

\paragraph{Invariants and well-formedness properties.} We present the
well-formedness properties we preserve over expressions
(\coqe{wf_expr}) and over hash-consing states (\coqe{wf_hashcons}) in
\figref{fig:pure-hc-inv}.
The inductive predicate \coqe{wf_expr st v e} depends on a variable
level \coqe{v} that indicates a bound over the variable identifiers
appearing in the BDD \coqe{e}.  This ensures that variable are kept
ordered.
Then, a well-formed expression is either a leaf (cases \coqe{wf_T} and
\coqe{wf_F}) or a pointer to a hash-consing node with a head variable
\coqe{w} less than the bound \coqe{v}.
(Remark that this definition of well-formed expression is not directly
recursive, because the actual recursion takes place in the definition
of \coqe{wf_hashcons} below; this notion of well-formed expressions coupled
with well-formedness of the state suffice to prove that expressions
are DAGs with a suitable ordering on the variables.)

The predicate \coqe{wf_hashcons} is the general well-formedness
property over hash-consing states.
The predicate \coqe{wf_bijection} ensures that the two maps
\coqe{hmap} and \coqe{graph} form a bijection;
\coqe{wf_expr_lt_next} ensures that the next fresh variable counter
will indeed produce fresh variables;
\coqe{wf_map_wf_expr_l} and \coqe{wf_map_wf_expr_r} ensure that a node
stored in the hash-consing structure has well-formed children,
respecting the variable order;
\coqe{wf_reduced} ensures that all those nodes are reduced (ie. their
left child is different from their right child).

Note that the statement that the hash-consing structure is correct
corresponds to \coqe{wf_bijection} and \coqe{wf_expr_lt_next}. The
other statements correspond to properties about BDDs, namely the
facts that they are reduced and ordered.

\begin{figure}[t]
\centering
\begin{coq}
Inductive wf_expr st : var -> expr -> Prop :=
| wfe_T : forall v, wf_expr st v T
| wfe_F : forall v, wf_expr st v F
| wfe_N : forall (p : positive) (l : expr) (w: var) (h : expr),
            get p st.(graph) = Some (l, w, h) ->
            forall (v : var), (w < v) ->
            wf_expr st v (N p).

Record wf_hashcons (st : hashcons_state) : Prop := {
  wf_bijection : forall p n, get n st.(hmap) = Some p <-> get p st.(graph) = Some n;
  wf_expr_lt_next : forall p v, wf_expr st v (N p) -> p < st.(next);
  wf_map_wf_expr_l : forall p x v y, get p st.(graph) = Some (x, v, y) -> wf_expr st v x;
  wf_map_wf_expr_h : forall p x v y, get p st.(graph) = Some (x, v, y) -> wf_expr st v y;
  wf_reduced : forall p l v h, get p st.(graph) = Some (l, v, h) -> l <> h
}.
\end{coq}
\caption{Well-formedness of hash-consing state}\label{fig:pure-hc-inv}
\end{figure}

\paragraph{The denotation of BDD expressions.} Recall that BDD
expression, as we have defined them in this section, do not have an
inductive structure we can follow: this means that we cannot
define the semantics of BDD expressions as a Coq fixed-point.
Rather, we have to define the semantics of BDD expressions as an
inductive predicate that uses the state of the hash-consing
data structure to go through the graph.
The inductive \coqe{value} is shown on \figref{fig:pure-value}. It is
defined as a binary relation with two parameters (the valuation of
variables \coqe{env} and the state of the hash-consing structure
\coqe{st}) and two arguments: the expression and its denotation (a
Boolean).

\begin{practical}
  We only use the type of environments \coqe{env} to establish a
  semantics, thus efficiency is not an issue. However, we prefer to
  use finite maps of some kinds to represent environments: using a
  function of type \coqe{var -> bool} would force the client to reason
  about the fact that some BDD expressions are insensitive to the
  valuation of some variables. Using a finite map make this reasoning
  internally in our library and eases the client's life.
\end{practical}

\begin{figure}
  \centering
\begin{coq}
Inductive value env st : expr -> bool -> Prop :=
| value_T : value env st T true
| value_F : value env st F false
| value_N : forall (p : positive) (l : expr) (v : var) (h : expr),
              get p st.(graph) = Some (l, v, h) ->
              forall (vv : bool), get env v = Some vv ->
              forall (vhl : bool), value env st (if vv then h else l) vhl ->
                                 value env st (N p) vhl.
\end{coq}
\caption{The denotation of BDD expressions}\label{fig:pure-value}
\end{figure}

\paragraph{Termination of BDD operations.}
The first problem that we have to solve in our Coq representation is
that, as can be expected from our data structure, BDD operations cannot be
defined using structural recursion (there is no inductive structure to follow).
Unfortunately, we cannot easily use well-founded recursion here because
the well-founded relation involves both parameters of the function and
the global state.

The problem is that the termination of the BDD operations relies on
the fact that the graph of nodes is acyclic; but the graph is not
fixed through an execution of the melding operation! Rather, the
global state is threaded in the recursive calls of $\meld$. Therefore,
proving the recursive calls to be well-founded requires to prove that
the $\meld$ operation is monotonic w.r.t. graph inclusion. In order to
prove termination, we would have to define the fixed-point and prove this
monotonicity property \emph{at the same time}. This would involve
embedding invariants directly in the global state, using dependent
types. This has two major drawbacks: first, defining terms containing
complex dependent types is cumbersome and proving properties about
them is often very challenging. Second, we would have to pay close
attention to prevent Coq from normalizing those proof terms if we want
to use this library for reflection purposes.

In the end, we resorted to define partial functions that use a
\emph{fuel} argument to ensure termination: that is, they use an
explicit bound on the number of iterations to do.

% A typical fixpoint combinator defined using fuel is shown on
% \figref{fig:fuel-fixpoint}. It approximates a fixpoint combinator by
% limiting the maximal depth of recursion of a function \coqe{f}.

% \begin{figure}
%   \centering
% \begin{coq}
% Section fuel_fix.
%   Context {A : Type}.
%   Variable P : A -> Type.

%   Fixpoint fuel_fix (f : (forall y : A, option (P y)) forall x : A, option (P x) ) n :=
%   match n with
%       0 => fun x => None
%     | S n => fun x => f (fuel_fix f n) x
%   end.
% End fuel_fix.
% \end{coq}
% \caption{A typical fixpoint combinator defined using fuel}\label{fig:fuel-fixpoint}
% \end{figure}
% Small variations on this very same concept make it possible to iterate
% a function lazily $2^n$ times, but are out of the scope of this
% paper. \jh{Not out of the scope?}

\paragraph{Memoizing operations.}
Finally, it is possible to enrich our hash-consing structure with
memoization tables in order to tabulate the results of BDD
operations.
\begin{figure}
  \centering
\begin{twolistings}
\begin{coq}
Record memo := {
  mand : (positive * positive) $\fmap$ expr;
  mor  : (positive * positive) $\fmap$ expr;
  mxor : (positive * positive) $\fmap$ expr;
  mneg : positive $\fmap$ expr}.
\end{coq}
&
\begin{coq}
Record state := {
  ... :> hashcons_state;
  ... :> memo}.

$ $
\end{coq}
\end{twolistings}
  \caption{Adding memoization tables to the global state}\label{fig:pure-bdd-memo}
\end{figure}

The memoization tables (see \figref{fig:pure-bdd-memo}) are passed
around by the state monad, just as the hash-consing structure. It is
then necessary to maintain invariants on the memoization
information.

\begin{practical}
  In the definition of state, we use two \emph{field coercions}
  (denoted \coqe{:>}) to declare coercions from the type \coqe{state}
  to the types \coqe{hashcons_state} and \coqe{memo}. In practice,
  this means that one can use a \coqe{state} in any expression that
  would expect the result of one of these two projections.
\end{practical}

\begin{figure}
  \centering
\begin{coq}
Record wf_memo2 (st : hashcons_state) (m : (positive * positive) $\fmap$ expr)
                 (opb : bool -> bool -> bool)
:= {
  wf_memo2_find_wf_res :
    forall x y e, get (x, y) m = Some e -> wf_expr st v (N x) -> wf_expr st v (N y) ->
      wf_expr st v e;
  wf_memo2_find_wf :
    forall x y e, get (x, y) m = Some e -> exists v, wf_expr st v (N x) /\ wf_expr st v (N y);
  wf_memo2_find_sem :
    forall x y res, get (x, y) m = Some res ->
      forall env vx vy, value env st (N x) vx -> value env st (N y) vy ->
        value env st res (opb va vb)
}.

Record wf_memo_neg (st : hashcons_state) (m : positive $\fmap$ expr) := { ... }.

Record wf_memo (st : state) := {
  wf_memo_mand : wf_memo2 st st.(mand) Datatypes.andb;
  wf_memo_mor  : wf_memo2 st st.(mor) Datatypes.orb;
  wf_memo_mxor : wf_memo2 st st.(mxor) Datatypes.xorb;
  wf_memo_mneg : wf_memo_neg st st.(mneg)
}.

Record wf_st (st : state) : Prop := { ... : wf_hashcons st; ... : wf_memo b }.
\end{coq}
\caption{Invariant over the memoization information}\label{fig:pure-hc-inv-memo}
\end{figure}

We present the invariants over the memoization maps for the binary
operations\footnote{The case of the negation operation is similar and is not detailed.} in
\figref{fig:pure-hc-inv-memo}.
First, we have to prove that the nodes referenced in the domain and in
the codomain of those tables are well-formed and that those tables
keep the bounds over the variables correct. Then, we have to state
that the memoization information is semantically correct.
One downside of this data structure definition is that we are forced
to define one table per operation that we want to memoize, and that
this is not modular: adding a new operation requires modifying the
definition of the memoization state and add the corresponding field in
the \coqe{wf_memo} record.

(Note that finding the correct pattern of memoization for a program is
still an art rather than a science: using the data structure above, we
keep the memoized values from one run of an operation to the other. In
this section, we will settle on this conservative strategy, but other
memoization strategies are possible that yield different performance
and memory consumption profiles.)

\paragraph{A mouthful of code.}
The final version of our code is shown on
\figref{fig:pure-combinator}. We use a do-notation \`a la Haskell to
make it more palatable.

\begin{figure}
  \centering
\begin{coq}
Section combinator.
  (* fx is the base case when one operand is F *)
  Variable fx : expr -> state -> option (expr * state).
  (* tx is the base case when one operand is T *)
  Variable tx : expr -> state -> option (expr * state).
  Variable memo_get : positive -> positive -> state -> option expr.
  Variable memo_update : positive -> positive -> expr -> state -> state.

  Definition memo_node a b l v h st :=
    let (res, st) := mk_node l v h st in
    let st := memo_update a b res st in
    (res,st).

  Fixpoint combinator depth : expr -> expr -> state -> option (expr * state) :=
    fun a b st =>
    match depth with
    | 0 => None
    | S depth =>
      match a,b with
        | F, _ => fx b st
        | _, F => fx a st
        | T, _ => tx b st
        | _, T => tx a st
        | N na, N nb =>
          match memo_get na nb st with
            | Some p => Some (p,st)
            | None =>
              do nodea <- get na st.(graph);
              do nodeb <- get nb st.(graph);
              let '(l1,v1,h1) := nodea in
              let '(l2,v2,h2) := nodeb in
              match Pos.compare v1  v2 with
                | Eq =>
                  do x, st <- combinator depth l1 l2 st;
                  do y, st <- combinator depth h1 h2 st;
                  Some (memo_node na nb x v1 y st)
                | Gt =>
                  do x, st <- combinator depth l1 b st;
                  do y, st <- combinator depth h1 b st;
                  Some (memo_node na nb x v1 y st)
                | Lt =>
                  do x, st <- combinator depth a l2 st;
                  do y, st <- combinator depth a h2 st;
                  Some (memo_node na nb x v2 y st)
              end
          end
      end
     end.
End combinator.
  \end{coq}
  \caption{Deep combinator for binary operations}
  \label{fig:pure-combinator}
\end{figure}

Then, we prove that under some hypotheses, this combinator is correct:
that is, it produces well-formed hash-consing structures and
memoization tables, and the denotation of the resulting expression
is correct.
For the sake of clarity, we will not expose these hypotheses
nor the resulting correctness theorem in this paper, and refer the
interested reader to the archive of code that accompanies this
paper~\cite{online-doc}.

\paragraph{Implementing the and operation.} However, we demonstrate
the use of this combinator on the particular example of the {\tt and}
function; all other binary operations follow the same pattern.
First, we have to define a function \coqe{upd_and} that updates the
memoization state: it is simply a wrapper that add an element to the
right memoization table, and leave the others untouched. Then, we
define the function \coqe{mk_and} as a simple call to the binary
combinator. The crux here is the choice of the functions \coqe{Fx} and
\coqe{Tx} that specify the behavior of the combinator at the leaves of
the DAG.

\medskip

\begin{twolistings}
  \begin{coq}
Definition upd_and na nb res (st : state) :=
  mk_state st
  {| mand := set (na,nb) res st.(mand);
     mor  := st.(mor);
     mxor := st.(mxor);
     mneg := st.(mneg)
  |}).
\end{coq}
&
\begin{coq}
Definition mk_and :=
  combinator
    (fun x st => Some (F,st) )  (* Fx *)
    (fun x st => Some (x,st) )  (* Tx *)
    (fun a b st => get (a,b) st.(mand))
    upd_and.
$ $
\end{coq}
\end{twolistings}

Then, the semantic correctness theorem for this operation is defined
as follows.
\begin{coq}
Theorem mk_and_correct depth v0 (st: state) a b :
  wf_st st ->
  wf_expr st v0 a -> wf_expr st v0 b ->
  forall res st', mk_and depth a b st = Some (res, st') ->
    wf_st st' /\ incr st st' /\
    wf_expr st' v0 res /\
    forall env va vb,
      value env st a va ->
      value env st b vb ->
      value env st' res (va && vb).
\end{coq}

This statement proves three things under the hypotheses that the given
state and BDDs are well-formed : first, \coqe{mk_and} returns a well
formed state. Second, it manipulates the state in an increasing manner
(and, in particular, previously well-defined BDDs will still be
well-defined and keep their semantic values). Third, the returned BDD
have the expected semantic interpretations.

\paragraph{Canonicity.} We can prove that this representation of BDDs
is canonical: that is, well-formed equivalent expressions are mapped
to the same nodes.
\begin{coq}
Definition equiv st e1 e2 :=
  forall env v1 v2, value env st e1 v1 ->  value env st e2 v2 -> v1 = v2.

Lemma canonicity st v e1 e2 :
  wf_st st -> wf_expr st v e1 -> wf_expr st v e2 ->
  equiv st e1 e2 -> e1 = e2.
\end{coq}

From this result, it follows that the (non-recursive) \coqe{eqb}
function from \figref{fig:pure-hashcons} is a correct and complete
characterization of semantic equivalence of expressions.
\begin{coq}
Lemma eqb_correct st v e1 e2 :
  wf_st st -> wf_expr st v e1 -> wf_expr st v e2 ->
  (eqb e1 e2 = true <-> equiv st e1 e2).
\end{coq}

\paragraph{Garbage collection.} In the above version of BDDs, we have
not implemented garbage collection. That is, allocated nodes are never
destroyed, until the allocation map becomes unreachable as a
whole. Garbage collection could be added, e.g., using a stop and copy
operation that preserve a set of roots. This is beyond the scope of
this paper.

\subsection{The {\sc pure-shallow} approach}\label{sec:pure-shallow}
The previous implementation uses a \emph{deep embedding} of the
representation of the BDD in memory via the \coqe{graph} map. This is
a natural way to encode a directed acyclic graph, but, as we saw,
makes it difficult (if not unfeasible) to deal properly with
termination. Therefore, we would like to be able to reason about BDDs
as if they formed an inductive type, while keeping the ability to
share sub-terms at runtime.

There is actually no need to look further than inductive types to
do that. The standard intuition about inductive types is that they
define the smallest type closed under application of constructors: the
mental image that we get from that is a \emph{tree}. Yet, there is
nothing that prevent us to use the system to share sub-terms.%
\footnote{Such an approach may be interesting even if each structure
  is a tree, but different trees can share sub-trees. For instance,
  the Astr\'ee static analyzer implements maps as balanced binary
  trees with \emph{opportunistic}
  sharing~\cite[\S6.2]{BlanchetCousotEtAl02-NJ}: e.g. when an
  idempotent operation $f(x,x)=x$ is to be performed on the images of
  two maps $g$ and $h$, returning the map $x \mapsto f(g(x),h(x))$,
  then if $g$ and $h$ are determined to be identical through pointer
  equality, $g$ is directly returned.
  % Thomas: Reviewer 1 commented on the following sentence. Since I am
  % not quite sure about what it means either, I remove it.
  % Thus, applying such a function to two maps that differ only on few
  % values is considerably faster than traversing the two maps.
}

In this section, we demonstrate that we can encode BDDs in Coq using a
representation that looks like binary decision \emph{trees}, yet has
runtime performances similar to the \textsc{pure-deep} implementation
(see \secref{sec:pure-deep}), using explicit sharing. We present on
\figref{fig:shallow} our inductive definitions, and the associated
allocation function \coqe{mk_node}. This has to be compared with
\figref{fig:pure-hashcons}: the difference mainly lies in the deletion
of the \coqe{graph} map, and the explicit recursive structure of the
\coqe{expr} type.

\begin{practical}
  The reader may wonder why we chose to define \coqe{expr} as two
  mutual inductive types \coqe{expr} and \coqe{hc_expr}.
  Indeed, we explain in \secref{lambda-shallow} that it is possible to
  inline \coqe{hc_expr} inside the definition of \coqe{expr}.
  The mutual inductive solution is inspired by the hash-consing
  library in OCaml by Conchon and
  Filli\^atre~\cite{ConchonFilliatre06wml}.  This library makes a
  clear distinction between hash-consing nodes (the equivalent of our
  \coqe{hc_expr} inductive) and the actual values that are
  hash-consed. In Coq, this makes some inductive proofs a little bit
  more involved: we need to use mutual induction on the two
  data-types.
\end{practical}

\begin{figure}
  \centering
\begin{coq}
Inductive expr := | F | T | N : hc_expr -> var -> hc_expr -> expr
with hc_expr := HC : expr -> positive -> hc_expr.

(* Two extra definitions that are used as coercions in the following code*)
Definition unhc (e : hc_expr) := let 'HC res _ := e in res.
Definition ident (e : hc_expr) := let 'HC _ res := e in res.

Definition eqb a b :=
  match a,b with
    | T,T | F,F => true
    | N (HC _ id1) x (HC _ id2), N (HC _ id1') x' (HC _ id2') =>
      (id1 =? id1') && (x =? x') && (id2 =? id2')
    | _, _ => false
  end.

Record hashcons_state := {hmap : expr $\fmap$ hc_expr; next : positive}.

Definition alloc u st :=
  let r := HC u b.(next) in
  (r, {| hmap := set u r b.(hmap);  next := b.(next) + 1|}).

Definition hc_node (e : expr) (st : hashcons_state) :=
  match get e st.(hmap) with
    | Some x => (x, st)
    | None => alloc e st
  end.

Definition mk_node (l : hc_expr) (v : var) (h : hc_expr) st :=
  if (ident l =? ident h) then (l,st) else hc_node (N l v h) st.
\end{coq}
\caption{A shallow-embedding of sharing}\label{fig:shallow}
\end{figure}

The code of the \coqe{hc_node} function is subtle: a call to
\coqe{hc_node e bdd} will perform a lookup in the hash-consing map
\coqe{hmap}: if the same expression was previously allocated, then we
return the old version; otherwise, we update the map \coqe{hmap} with
a mapping from the expression to its hash-consed version.
The lookup ensures that equal expressions (modulo the comparison
function used to index \coqe{hmap}) are mapped to the same
hash-consed expression.
Then, \coqe{mk_node l v h bdd} will first test the identifiers of
\coqe{l} and \coqe{h} for equality: if it is the case, then there is
no need to introduce a new node; otherwise, we perform a call to
\coqe{hc_node}.

As an example, assume that \coqe{x} and \coqe{y} are expressions with
different identifiers. The following code
\begin{coq}
let (a,st) := mk_node x v y st in
let (b,st) := mk_node x v y st in
let (c,st) := mk_node a v' b st in ...
\end{coq}
will make \coqe{a}, \coqe{b} and \coqe{c} point to the same memory
location!  However, if \coqe{x} and \coqe{y} were not shared
maximally, then neither are \coqe{a}, \coqe{b} nor \coqe{c}.

\paragraph{A word on memoization.} There is no difference at all in
the way we handle memoization in this implementation
w.r.t. \secref{sec:pure-deep}. That is, we implement the same \coqe{state}
record as before; and pass the same memoization tables around.

\paragraph{A word on termination.} It is now easier to define
recursive functions that operate on BDDs by taking advantage of the
inductive definition of \coqe{expr}. We have to stress that
\emph{easier} is not \emph{easy} because the Coq termination checker
requires that recursive calls are made on a structurally smaller
argument: there is no built-in support for recursive definitions with
pairs of arguments that are decreasing w.r.t. a lexicographic
order. Therefore, we have to use nested fixed-points or prove that the
recursive calls are well-founded. In this section, we choose the
former.

\paragraph{Re-implementing the combinator.} We are now ready to
describe the code of the implementation of the binary combinator
presented in \figref{fig:shallow-combinator}. The code is similar to
the one in \figref{fig:pure-combinator} with a few key differences:
thanks to the inductive definition of expressions, we do not have to
perform lookups in the \coqe{graph} map and we do not use fuel
anymore. This makes the combinator function \emph{total}; and we can
get rid of the Maybe monad.

\begin{figure}
  \centering
\begin{coq}
Section combinator.
  Variable fx : hc_expr -> state -> hc_expr * state.
  Variable tx : hc_expr -> state -> hc_expr * state.
  Variable memo_get : positive -> positive -> state -> option (hc_expr).
  Variable memo_update : positive -> positive -> hc_expr -> state ->  state.

  Fixpoint combinator : hc_expr -> hc_expr -> state -> hc_expr * state := fun a =>
    fix combinator_rec b st :=
    match unhc a, unhc b with
      | F, _ => fx b st
      | _, F => fx a st
      | T, _ => tx b st
      | _, T => tx a st
      | N l1 v1 h1, N l2 v2 h2 =>
        let ida := ident a in
        let idb := ident b in
        match memo_get ida idb st with
          | Some p => (p,st)
          | None =>
            let '(res, st) :=
              match Pos.compare v1 v2 with
               | Eq =>
                  let '(x, st) := combinator l1 l2 st in
                  let '(y, st) := combinator h1 h2 st in
                  mk_node x v1 y st
                | Gt =>
                  let '(x, st) := combinator l1 b st in
                  let '(y, st) := combinator h1 b st in
                  mk_node x v1 y st
                | Lt =>
                  let '(x, st) := combinator_rec l2 st in
                  let '(y, st) := combinator_rec h2 st in
                  mk_node x v2 y st
              end
            in
            let '(_, st) := memo_update ida idb res st in
            (res, st)
        end
    end.
End combinator.
\end{coq}
  \caption{Shallow combinator for binary operations}
  \label{fig:shallow-combinator}
\end{figure}

\paragraph{Canonicity.} Again, we prove that this representation of
BDDs is canonical: well-formed equivalent expressions are mapped to
the same nodes. Again, we have the corollary that the (non-recursive)
equality test from \figref{fig:shallow} that inspects the (top-level)
node identifiers is a complete and correct characterization of semantic
equivalence.
\begin{coq}
Definition equiv e1 e2 :=
  forall env v1 v2, value env e1 v1 -> value env e2 v2 -> v1 = v2.

Lemma eqb_correct st e1 e2 v :
  wf_st st -> wf_expr st v e1 -> wf_expr st v e2 ->
  (eqb e1 e2 = true <-> equiv e1 e2).
\end{coq}

\paragraph{Comparison with the previous approach.} The implementation
presented in this section is derived from the previous one, with the
following improvements: the proofs are roughly 20\% shorter; the
performances are slightly better when executing the code inside Coq
(there is less administrative book-keeping to do in our
data structures); the functions that operates on BDDs are total.
Furthermore, it would probably be easier to implement garbage
collection in this setting than in the previous one, thanks to the
simpler definition of the global state.

The situation is almost ideal for the equality test: we prove that
the (non-recursive) equality function that inspects the top-level
identifiers of nodes is a correct and complete characterization of
semantics equivalence of BDD expressions. However, we have no way to
prove that it corresponds to physical equality. Actually, we cannot
\emph{state} within Coq that it is never the case that two identical
representations of the same term coexists, even if we could argue at a
meta-level that it is indeed not the case.

\section{From pure data structures to persistent data structures via extraction}
\label{sec:impure}
In the previous section, we use a state monad to store information
about hash-consing and memoization. However, one can see that, even if
these programming constructs use a mutable state, they behave
transparently with respect to the pure Coq definitions.

We have seen earlier (see \secref{ssec:adjustable-references}) that if
we abandon (efficient) executability inside Coq, we can express new
idioms.
In the following, we implement the BDD library in Coq as if
manipulating decision trees with neither sharing, nor hash-consing
tables, nor memoization tables, then add the hash-consing and
memoization code by using special features of the extraction mechanism
to remap some constants arbitrarily to custom OCaml code.

\subsection{The \textsc{smart} approach}\label{sec:smart}
% An additional benefit is that, since we use native hash tables, we may
% as well use \emph{weak} ones, enabling the native garbage collector to
% reclaim unused nodes without being prevented from doing so by the
% pointer from the table.

More precisely, we define our BDDs as binary decision trees (see
\figref{fig:coq-smart-hashcons}), and implement operations in Coq on
this simple data structure.
Then, we tell Coq to extract the \coqe{bdd} inductive type to a custom
\ocamle{bdd} OCaml type (see left of
\figref{fig:ocaml-smart-hashcons}) and to extract constructors into
\emph{smart constructors} that maintain the maximal sharing
property. The type defined in OCaml is identical to the Coq one,
except that it carries one extra field of type \ocamle{tag}, morally
containing the associated unique identifier.
The smart constructors make use of the hash-consing library used in
Why3~\cite{Why3}, a recent version of a library by Conchon and
Filli\^atre~\cite{ConchonFilliatre06wml}. It defines the
\ocamle{Hashcons.Make} functor, that we instantiate. The generated
module provides a \ocamle{HCbdd.hashcons} function that returns a
unique hash-consed representative for the parameter.

\begin{figure}[t]
\begin{subfigure}[t]{\textwidth}
  \centering
\begin{twolistings}
\begin{coq}
Inductive bdd: Type :=
| T | F | N : var -> bdd -> bdd -> bdd.
\end{coq}
&
\begin{coq}
Extract Inductive bdd =>
  "bdd" ["hT" "hF" "hN"] "bdd_match".
\end{coq}
\end{twolistings}
  \caption{BDDs in Coq as decision trees}\label{fig:coq-smart-hashcons}
\end{subfigure}

\begin{subfigure}[t]{\textwidth}
  \centering
\begin{twolistings}
\begin{ocaml}
type bdd =
| T of tag | F of tag | N of positive * bdd * bdd * tag

module HCbdd = Hashcons.Make(...)

let hT = HCbdd.hashcons (T Weakhtbl.dummy_tag)
let hF = HCbdd.hashcons (F Weakhtbl.dummy_tag)
let hN (v, t, f) = HCbdd.hashcons (N (v, t, f, Weakhtbl.dummy_tag))
\end{ocaml}
&
\begin{ocaml}
let bdd_match fT fF fN t =
  match t with
  | T _ -> fT ()
  | F _ -> fF ()
  | N (v, t, f, _) -> fN v t f
\end{ocaml}
\end{twolistings}
  \caption{Hash-consed OCaml BDD type}\label{fig:ocaml-smart-hashcons}
\end{subfigure}
\caption{Implementing BDDs in Coq, extracting them using smart
  constructors}
\end{figure}

The reader may notice that we choose to name \coqe{bdd} in Coq what is
clearly a representation of a binary decision tree, and which
corresponds to what was previously named \coqe{expr}.
We believe that this particular choice of name makes sense if we
consider values of type \coqe{bdd} to represent Boolean functions.

\paragraph{Discussion: the status of the equality test.}
In Coq, we define the obvious recursive function \coqe{bdd_eqb} of type
\mbox{\coqe{bdd -> bdd -> bool},} that decides structural equality of
BDDs.
Then, we extract this function into OCaml's physical equality:

\begin{coq}
Fixpoint bdd_eqb (b1 b2:bdd): bool :=
  match b1, b2 with
    | T, T | F, F => true
    | N v1 b1t b1f, N v2 b2t b2f =>
      Pos.eqb v1 v2 && bdd_eqb b1t b2t && bdd_eqb b1f b2f
    | _, _ => false
  end.

Extract Inlined Constant bdd_eqb => "(==)".
\end{coq}

From a meta-level perspective, we argue that the two are equivalent
thanks to the physical unicity of hash-consed structures, provided
that all values are constructed using our smart-constructors, which is
the case if we create all nodes from code extracted from Coq (of
course, handwritten OCaml code may break this invariant). The key
point here is that the way we build terms enforces the fact that the
BDDs they build are maximally shared.

\paragraph{Keeping the BDD reduced.}
One could be tempted to put in the OCaml code of the smart constructor
\ocamle{hN} a test that would enforce that BDDs are reduced. That is,
it would not build a node if its two children were identical.
\begin{ocaml}
let hN (v,t,f) =
  if t == f
  then t
  else HCbdd.hashcons (N (v, t, f, Weakhtbl.dummy_tag))
\end{ocaml}
To see the problem with this idea, consider the following Coq code.
\begin{coq}
match N (v,T,T) with
| T => false
| N (_,_,_) => true
end
\end{coq}
In Coq, the above expression reduces to \coqe{true} while the
extracted version would reduce to \coqe{false}.
Therefore, this idea is wrong: avoiding the node construction makes
subsequent case analysis behave inconsistently between Coq and OCaml.

In the end, the correct way to implement reduction is to use the
following helper function, written in Coq, that builds a node only
when necessary.
\begin{coq}
Definition N_check (v : var) (bt bf : bdd) : bdd :=
  if eqb bt bf then bt else N v bt bf.
\end{coq}
Note that the user is forced to use this function instead of using the
\coqe{N} constructor\footnote{Still, the \coqe{N} constructor that
  occurs in the definition \coqe{N_check} must be extracted using a
  smart-constructor.}.
Failing to do so will result in code that is correct, but does not
build reduced binary decision diagrams. More precisely, extracting
\ocamle{eqb} to \ocamle{==} would make it possible to prove that BDD
operations are semantically correct, but this would make diagrams
non-canonical: well-formed expressions could be represented by several
different nodes.

\paragraph{Implementing the combinator.}
The last ingredient needed to transform a decision tree library into a
BDD library is memoization. We use the same kind of ideas: we define
our functions as if not memoized, but we use a special well-founded
fixed-point combinator, that we extract to a memoizing fixed-point
combinator. The details can be seen on \figref{fig:memoization}: we
declare an abstract type class \coqe{memoizer} of types \coqe{A} such
that we know how to memoize functions of type \coqe{forall x : A, P x}.
This is extracted in OCaml to the type of polymorphic fixed-point
combinators, with an extra technical \ocamle{int} parameter used to
specify the initial size of the used hash map. In Coq, we then define
a memoizing combinator \coqe{memo} and a memoizing fixed-point combinator
\coqe{memo_rec} as if they were not using memoization, but we ask the
extraction mechanism to map them to special OCaml functions, that make
use of the type class instance given as parameter. These functions are
observationally equivalent to the Coq ones, provided that the type
class instance is correct, and that the memoized function is pure.

\begin{figure}
  \begin{subfigure}{\textwidth}
  \begin{coq}
Parameter memoizer : forall A : Type, Type.
Existing Class memoizer.
Extract Constant memoizer "'key" => "'key Helpers_common.memoizer".

Definition memo A {H : memoizer A} P := @id (forall x : A, P x).
Extract Inlined Constant memo => "Helpers_common.memo".
Arguments memo [A] {H} [P] _ _. (* Set implicit arguments for memo*)

Definition memo_rec A {H : memoizer A} := @Coq.Wf.Fix A.
Extract Inlined Constant memo_rec => "Helpers_common.memo_rec".
Arguments memo_rec [A] {H} [R] Rwf [P] F x. (* Set implicit arguments for memo_rec *)
  \end{coq}
  \caption{Coq part}
  \end{subfigure}
  \begin{subfigure}{\textwidth}
    \begin{ocaml}}
type 'key memoizer =
    { memo : 'a. int -> (('key -> 'a) -> ('key -> 'a)) -> ('key -> 'a) }

let memo m f = m.memo 5 (fun _ x -> f x)

let memo_rec m f = m.memo 5 (fun frec x -> f x (fun y _ -> frec y))
\end{ocaml}
  \caption{OCaml part}
  \end{subfigure}
  \caption{Memoizing combinators}
  \label{fig:memoization}
\end{figure}

It is worth noting that directly exposing the type \ocamle{'key memoizer}
in Coq would be unsound, because this allows to use its instances in a
non-terminating manner: the \coqe{memo_rec} wrapper makes sure the
recursive calls are well-founded. Moreover, it is important to
understand that we have not axiomatized these combinators. Instead, we
give real implementations, semantically equivalent to their OCaml
counterparts. This is important, because it then becomes clear that we
do not introduce any logical inconsistencies in Coq, and these terms
keep a computational content which could be very useful in proofs.

\begin{figure}
  \begin{subfigure}{\textwidth}
    \begin{coq}
Parameter memoizer_N : memoizer N.
Existing Instance memoizer_N.
Extract Inlined Constant memoizer_N => "Helpers_common.memoizer_N".
    \end{coq}
    \caption{Coq part}
  \end{subfigure}
  \begin{subfigure}{\textwidth}
    \begin{twolistings}
    \begin{ocaml}
module NHT =
  Hashtbl.Make (struct
    type t = coq_N
    let equal = (=)
    let hash = N.to_int
  end)
    \end{ocaml}
&
    \begin{ocaml}
let memoizer_N =
  { memo = fun n f ->
    let h = NHT.create n in
    let rec aux x =
      try NHT.find h x with Not_found ->
        let r = f aux x in NHT.replace h x r; r
    in aux }
    \end{ocaml}
    \end{twolistings}
    \caption{OCaml part}
  \end{subfigure}
  \caption{An instance of \coqe{memoizer} for \coqe{Coq.Numbers.BinNums.N}}
  \label{fig:memoization-N}
\end{figure}

This part of the code is modular, and can be used to memoize functions
of any domain. It is up to the user to give instances of the
\coqe{memoizer} type class: to do so, he can provide dedicated code,
as shown in \figref{fig:memoization-N} for the type \coqe{N} of
binary natural integers, using simple hash tables. Alternatively, for
the \coqe{bdd} type, our type class instance is just an OCaml wrapper
around the hash-consing library built-in memoization mechanisms.

\begin{figure}
\begin{coq}
Definition bdd_and : bdd -> bdd -> bdd.
refine
  (memo_rec (well_founded_ltof _ bdd_size) (fun x =>
     match x with
       | T => fun _ y => y
       | F => fun _ _ => F
       | N xv xt xf => fun recx =>
         memo_rec (well_founded_ltof _ bdd_size) (fun y =>
           match y with
             | T => fun _ => x
             | F => fun _ => F
             | N yv yt yf => fun recy =>
               match Pos.compare xv yv with
                 | Eq => N_check xv (recx xt _ yt) (recx xf _ yf)
                 | Lt => N_check yv (recy yt _) (recy yf _)
                 | Gt => N_check xv (recx xt _ y) (recx xf _ y)
               end
           end)
     end));
  unfold ltof; simpl; clear; abstract omega.
Defined.
\end{coq}
\caption{The and operation on bdds}
\label{fig:and-smart}
\end{figure}

Again, we use the example of the \coqe{bdd_and} operation, shown in
\figref{fig:and-smart}.
As in \figref{fig:shallow-combinator}, we define this function using
two nested fixed-points, in order to handle the special recursion scheme
of this function. The definition of \coqe{bdd_and} uses
\coqe{memo_rec} twice, in a nested fashion. This is needed in order to
handle the special recursion scheme of this function (decreasing on
one of its two parameters).

\paragraph{Garbage collection.} The strategy we use in
\secref{sec:pure-shallow} for garbage collection is very naive: we kept
everything alive, forbidding any garbage collection. Here, the
hash-consing library we use allow the garbage collector to reclaim any
node that is not referenced by the program, by using suitable weak
hash tables. Moreover, it reclaims any memoized value that is
associated with a dead key (we will not give more details here and
refer the reader to~\cite{ConchonFilliatre06wml}). While it avoids
memory leaks, this strategy does not necessarily give the best
performances (see \secref{sec:discussion}). We believe the changes
necessary to implement a new strategy are small, and do not involve
rewriting the proofs: only OCaml code is involved.

It is important to note that we have to memoize the functions each
argument after the other (keeping them curried). Indeed, one should be
tempted to memoize a function \coqe{bdd * bdd -> bdd}. This is not a
good idea, since we use weak hash-tables for the memoization: in this
case, the pairs containing the arguments are no longer accessible
after the function call, so that the garbage collector is going to
collect most of the memoized data at each cycle.  Using our pattern,
we make sure a memoized datum is not reclaimed as long as both
parameters are still alive in memory.
(Note that if we choose to use regular hash-tables instead of weak
ones, this argument does not stand anymore, but the program has a
different performance and memory consumption profile.)

\paragraph{Discussion: comparison with previous approaches.}
In this instance of our BDD library, all Coq definitions are kept
simple and proofs are straightforward. That is, we can prove semantic
correctness of all operations directly using structural induction on
decision trees and there is no state holding structures. There is a
nice separation between the hash-consing and memoization code, that is
generic, written in OCaml and not proved and the BDD code, mostly
written and proved in Coq. We discuss the lack of confidence in the
OCaml code and how it can be restored in \S\ref{ssec:trust_tweaked_extracted}.

We do not
need to use monads in the Coq code, so that the interface of the BDD
library remains modular and easy to use. Moreover, it is
straightforward to implement garbage collection strategies in order to
avoid memory leaks.

\subsection{The \textsc{smart+uid} approach}
\label{sec:smart+uid}
The previous \textsc{smart} approach totally hides the unique
identifiers from the Coq code. Yet, exposing these unique identifiers
may be useful at times.

Consider the following use case: from a BDD $B$ we would like to build
an equivalent propositional logic formula \emph{of linear size}, for
instance for feeding into a satisfiability modulo theory solver.  In
order to avoid an exponential blow-up, each shared sub-tree should
generate one single sub-formula, used in a ``let'' binder so that its
value can be used in multiple occurrences.  The obvious way to
implement such a transformation is to first detect which sub-trees are
shared, using a set of shared subtrees seen so far, then to perform
the transformation, using a table of mappings from subtrees to bound
variables.

It seems therefore highly desirable to be able to build sets and maps
over our hash-consed type. Generic functional sets and maps are
usually implemented using balanced trees over a totally ordered
data type; for efficiency, the comparison function should be very fast.
An obvious choice would be to expose the unique identifiers to the Coq
code (through a function \coqe{bdd -> uid}), or at least the total
order that they induce (through a function \coqe{bdd -> bdd -> comparison}).

Unfortunately, doing so without precautions can lead to unsoundness.
Consider a program where, in succession, two nodes $A$ and $B$ are
allocated, then a node $A'$ isomorphic to $A$ is created; let
$u_A < u_B$, $u_{A'}$ be the successive unique identifiers. If $A$ is
collected between the allocations of $B$ and $A'$, then $A'$ will be
allocated, with $u_{A'} > u_B$. Yet, $A'$ and $A$ are, from the point
of view of the Coq code, identical; thus $u_{A'}=u_A$, yielding an
inconsistency.

The workaround is to use a normal hash-table, as opposed to a weak
hash-table, which prevents the collection of unreachable nodes. Then,
two identical nodes created within the same execution are necessarily
physically equal and thus share the same identifier.

One difficulty remains. Gallina is a purely functional language;
the evaluation of a given term always yields the same result, and one
expects the same property to extend to the extracted OCaml program, as
long as it does not interact with the external world (e.g. reading
from files).  Yet, this is not necessarily the case if one exposes the
unique identifiers. Consider a program $P_1$, such that the extracted
OCaml code allocates two nodes $A$ and $B$ in this order. If $P_1$ is
run stand-alone, then $u_A < u_B$. Yet, if another program $P_2$
allocating $B'$ isomorphic to $B$ is first run, and $B'$ is not
collected, then $B'$ is the same as $B$ and $u_B=u_{B'}<u_A$.

It seems questionable that the result of an evaluation should depend
on whether or not some other (unrelated) evaluation has taken place,
if only because it makes debugging difficult.
\footnote{One could argue that, with certified programs in the Coq
  fashion, there is no need to debug: each function or module comes
  with a proof of its correctness, which compositionally entail the
  correctness of the whole program.  Yet, commonly one only proves the
  results to be \emph{correct}, not necessarily \emph{optimal}, and
  one proves very seldom that the computation has the expected
  complexity.  Furthermore, some computations are split between an
  untrusted solving procedure, and a trusted checker; a failed check
  entails having to debug the untrusted procedure, which may be hard
  if behaviors are hard to reproduce independently of the rest of
  program.}

\subsubsection{Maps and sets over hash-consed types}
Arguably, the final result of an algorithm using functional sets and
tables should not depend on the order relation used.  One could thus
opt to expose neither the unique identifiers, nor the order relation
they induce, but only a functional map interface (with \emph{find},
\emph{remove} and \emph{update}).  Then, the results cannot depend on
garbage collection and other disturbances to the ordering.  Yet, one
could also want these sets and maps to provide a \emph{fold}
operation; this would need to be restricted to associative and
commutative operations (so that their result do not depend on the
ordering).  Such an interface would fit the \emph{map-reduce} approach
to aggregate objects.

One additional advantage is that, with this approach, it is possible
to use weak hash tables and garbage collection, as long as the sets
and maps point to the nodes used as keys and thus prevent their
collection: the ordering relation is then exposed to the Coq code only
through equality testing, be it explicit or implicit (through the use
of set or maps).%
\footnote{It is of paramount importance that the sets and maps use the
  actual nodes as keys (ordered by the order on unique identifiers),
  not just the unique identifiers: this prevents scenarios such as
  creating an object $A$, inserting a key $u_A$ into some set $S$,
  collecting $A$, creating another object $A'$ isomorphic to $A$, but
  with a different key $u_{A'}$, which could distinguished from $A$ by
  membership in~$S$.  In a nutshell, it is crucial that the Coq
  program cannot test for equality of unique identifiers separated
  from their associated objects.}

While this line of work seems powerful, we did not pursue it: the many
axioms needed do not inspire confidence; some nontrivial
meta-theoretic reasoning about the sequential execution of OCaml code
extracted from Coq would be needed.

\subsubsection{Axioms}\label{sec:axioms}
\begin{figure}[t]
\begin{multicols}{2}
\begin{lstlisting}[language=Coq]
Axiom var : Set.

Axiom uid : Set. 
Axiom uid_eqb : uid -> uid -> bool.
Axiom uid_eq_correct: forall x y : uid,
    (uid_eqb x y =true) <-> x=y.

Inductive bdd : Set :=
| T | F 
| N : uid -> var -> bdd -> bdd -> bdd.

Axiom mkN : var -> bdd -> bdd -> bdd.

Axiom mkN_ok :
forall v : var, forall bt bf : bdd,
  exists id, mkN v bt bf = N id v bt bf.

Inductive valid : bdd -> Prop :=
| valid_T : valid T
| valid_F : valid F
| valid_N : forall var bt bf,
  (valid bt) -> (valid bf) ->
  (valid (mkN var bt bf)).

Axiom shallow_equal_ok :
  forall id1 id2 : uid,
  forall var1 var2 : var,	
  forall bt1 bf1 bt2 bf2 : bdd,
    valid (N id1 var1 bt1 bf1) ->
    valid (N id2 var2 bt2 bf2) ->
    id1 = id2 ->
    N id1 var1 bt1 bf1 =
    N id2 var2 bt2 bf2.
    
\end{lstlisting}
\end{multicols}
\caption{Axiomatization of equality using unique identifiers}\label{fig:axioms}
\end{figure}
In the \textsc{smart} approach, hash-consing and memoization are done
after the fact, and are completely transparent for the user; yet we
felt a need to break this transparency by exposing the unique
identifiers.
In the following, we instead make more explicit the hypotheses that we
make on the representation of BDDs.  That is, we make visible in the
inductive type of BDDs that each BDD node has a ``unique identifier''
field (see \figref{fig:axioms}) and we take the node construction
function as an axiom, which is implemented in OCaml.
Note that nothing prevents the \Coq{} program from creating new BDD
nodes without calling this function \coqe{mkN}. Yet, only objects
created by it (or copies thereof) satisfy the \coqe{valid} predicate;
we must declare another axiom stating that unique identifier equality
is equivalent to Coq's Leibniz equality \emph{for valid nodes}.
Then, we can use unique identifiers to check for equality.

This approach is close to the \textsc{smart} approach. It has one
advantage, the fact that unique identifiers are directly accessible
from the Coq code.  The previously mentioned limitation applies:
since unique identifiers are visible, one cannot use weak hash tables
and the OCaml garbage collector.

The use of axioms is debatable. On the one hand, the use of axioms somewhat
lowers the confidence we can give in the proofs, and they make the code
not executable within Coq. On the other hand, these axioms are actually
used implicitly when extracting Coq constructors to ``smart constructors'':
they correspond to the meta-theoretical statement that these constructors
behave as native Coq constructors. Thus, they make explicit some of the
magic done during extraction.

We could gain confidence about the fact that our axiomatization is
consistent, by using Coq's module system. That is, we could
parameterize our development in this section by a suitable interface
specifying the various functions that we axiomatized.
Then, we could instantiate this interface either with axioms, and
extract them to an efficient yet unproven implementation in OCaml; or,
we could instantiate them with a pure Coq version. This would prove
that this set of axioms is not unsound in Coq.
This general approach was used by Vafeiadis
\cite{adjustable-references} to prove that his interface for
adjustable references is sound, and we could do the same here. We have
yet to conduct this experiment.

\section{Discussion}\label{sec:discussion}
In this section, we compare our design patterns on various
aspects. Let's start with the easy ones:
\begin{description}
\item[Executability inside Coq.] The \textsc{pure-deep} and
  \textsc{pure-shallow} implementations can be executed inside Coq,
  and have decent performances. The \textsc{smart} approach
  can also be executed inside Coq, but has dreadful performance (when
  executed inside Coq, it uses binary decision trees).  The
  \textsc{smart+uid} approach cannot be executed inside Coq.
\item[Trust in the extracted code.] Unsurprisingly, the \textsc{smart}
  and the \textsc{smart+uid} approaches yield code that is harder to
  trust, while the \textsc{pure-deep} and \textsc{pure-shallow}
  approaches leave the extracted code pristine.
\item[Proof.] From a proof-effort perspective, the \textsc{smart}
  approach is by far the simplest.
  The \textsc{smart+uid} approach involves the burden of dealing with
  axioms. However, it makes it easier to trust that what is formally
  proven corresponds to the real behavior of the underlying runtime.
  By comparison, the \textsc{pure-deep} and \textsc{pure-shallow}
  approaches required considerably more proof-engineering in order to
  check the validity of invariants on the global state. Note however
  that our proof arguments are much simpler in the latter one.
\item[Garbage collection.]  Implementing (and proving correct) garbage
  collection for the \textsc{pure-deep} or \textsc{pure-shallow}
  approaches would require a substantial amount of work.  By contrast,
  the \textsc{smart} approach make it possible to use OCaml's garbage
  collector to reclaim unreachable nodes ``for free''.
\item[Operations] As we have shown, binary operations can be handled
  with a single parametric combinator. All our implementations use
  such a combinator to implement conjunction, disjunction,
  exclusive-or and so on. There is little work to do to add support
  for any binary operation we may have overlooked. The situation is
  more complicated when it comes to ternary operations (such as the
  if-then-else Boolean operation). Implementing it using the
  \textsc{smart} approach, and proving it correct requires around 80
  lines of code. It would require a non-trivial amount of work to
  implement it using our \textsc{pure-deep} or \textsc{pure-shallow}
  approaches, and we have not conducted this experiment yet. Other
  operations that are relevant in a BDD library are function
  composition and quantification. We have yet to implement these in
  any of our libraries. Quantification over one variable can be
  defined as a structural recursion over one BDDs. As such, we think
  it can be easily defined in all our frameworks. Unary composition
  (that is, substitution of one variable with another Boolean
  function) can be defined as a structural recursion over two BDDs. As
  a consequence, implementing it in the \textsc{smart} framework seems
  simple. However, this particular recursion scheme does not fit the
  combinator of the \textsc{pure-deep} or \textsc{pure-shallow}
  approaches, and some additional work would be needed in order to
  implement it using those approaches. Moreover, a non-negligible
  amount of work would be needed to support variable arity composition
  and quantification in all libraries.
\end{description}

\subsection{Performances of the extracted code}
We evaluate the performances of the OCaml code that is extracted from
our ``pure'' (see \secref{sec:pure}) and ``impure'' (see
\secref{sec:impure}) libraries, and we pit them against a reference
library implemented in OCaml~(available from J.C.~Filli\^atre's web
page).
This reference library does not keep the memoization table alive from
one execution of an operation to another: for instance, each time the
{\tt and} function is called, a new memoization hash-table is
allocated. Therefore, to make up for a fair benchmark with our
implementations, we modified this reference library to use a
memoization strategy closer to ours. This alternative reference
implementation is designed ``reference (conservative)'' in what
follows.

\begin{figure}
  \centering
  \begin{subfigure}{ 0.8 \linewidth}
    \includegraphics[width=\textwidth]{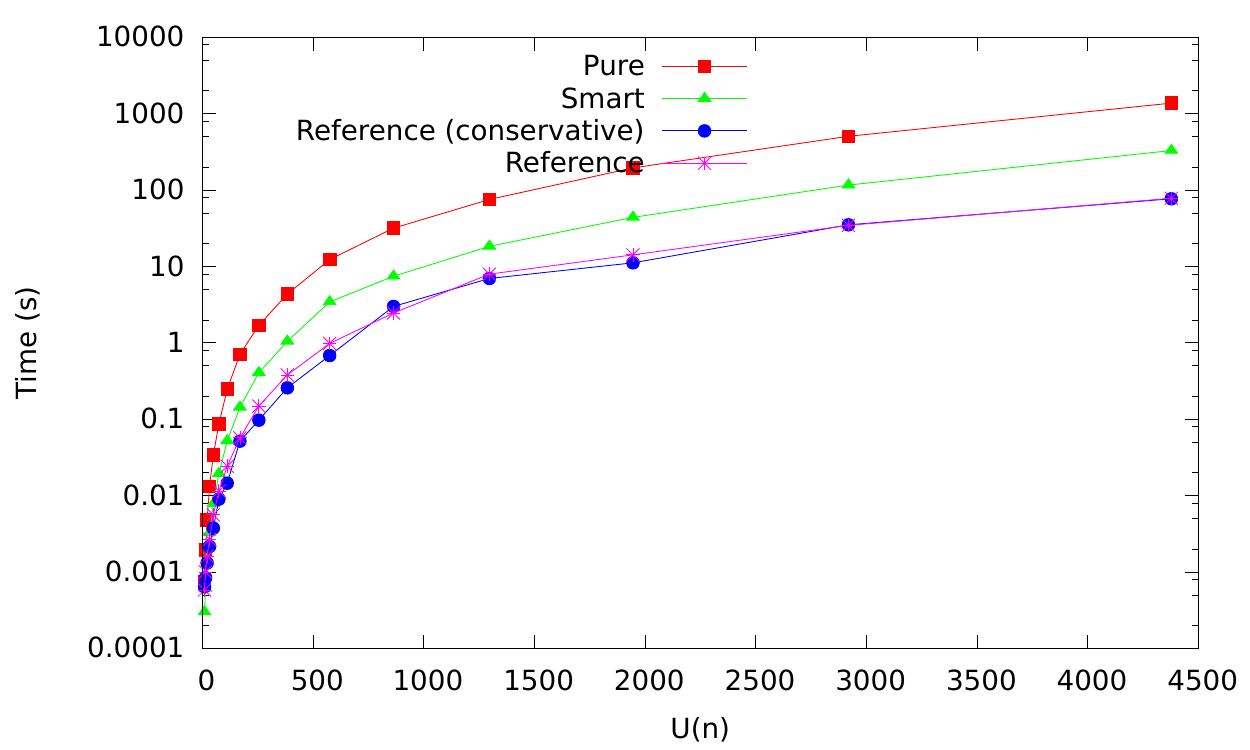}
    \caption{Urquhart's formula}
  \end{subfigure}

  \begin{subfigure}{ 0.8 \linewidth}
    \includegraphics[width=\textwidth]{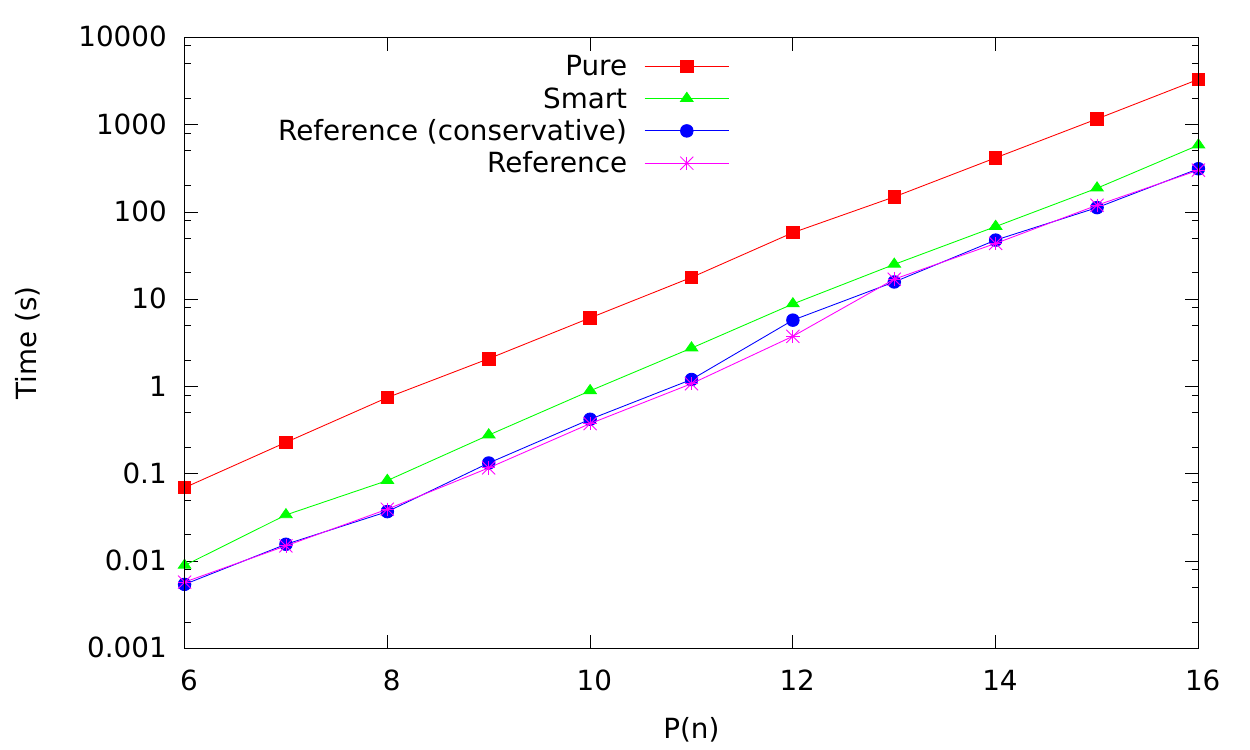}
    \caption{Pigeonhole principle}
  \end{subfigure}
  \caption{Execution time for the BDD benchmarks}
  \label{fig:benchmark}
\end{figure}

Then,  this evaluation, we use two standard benchmarks (see
\cite{DBLP:conf/asian/VermaGPA00}). The first one is Urquhart's
formula $U(n)$ defined by
\begin{displaymath}
U(n) \triangleq  x_1 \iff (x_2 \iff \dots (x_n \iff (x_1 \iff \dots (x_{n-1} \iff x_n))))
\end{displaymath}
The second kind of formula states the pigeonhole principle $P(n)$ for $n+1$
pigeons: that is, if there is $n+1$ pigeons in $n$ pigeonholes, then
there is at least one hole that contains two pigeons.
The plots of the execution time required to check that these formulas
are tautology are given in \figref{fig:benchmark}.

First, we remark that the reference implementation that we use is
(almost always\footnote{This is not necessarily true for small
  formulas due to differences in the start-up costs of these
  libraries.}) 4 times faster than our best implementation
\textsc{smart}. There is no significant difference in execution time
between the reference implementation with the conservative memoization
strategy and the reference implementation. We have not conducted a
detailed measurement of the memory consumption\footnote{These
  particular benchmarks were run on a computer with 1 TB of ram, which
  made it possible to avoid swapping during our experiments. Indeed,
  memory usage is the limiting factor on bigger instances of the
  tests.}, though, and the memory consumption profile is likely to be
different between the two.

Then, our \textsc{smart} implementation is roughly 4 times faster than
the \textsc{pure-deep} one: interestingly enough this value is
consistent over our benchmark, while we could have expected
logarithmic factors to show up. Indeed, recall that the pure
implementation uses functional finite maps, while the smart one uses
OCaml hash-maps.

Also, we have run the same benchmark on our \textsc{pure-shallow}
implementation, and the resulting plot is close to the
\textsc{pure-deep} plot: for small instances, the difference of
execution times are of about 10\%. For our largest instance of the
pigeonholes problem, we observe that the \textsc{pure-shallow} is
faster by about 30\%. We do not delve too much on this result: we
wonder for instance to what extent changing the implementation of
finite maps that we use could result in similar differences of
performances.

Primary investigation show that the main performance difference
between the reference implementation and our \textsc{smart} approach
come from the use of the generic hash-consing library: the reference
library uses a specialized hash-consing system, much faster than the
generic library we use. Similarly, the difference between the
\textsc{smart} and the \textsc{pure-shallow} seems to be the use of
AVL trees to represent dictionaries.

We did not investigate much the memory behavior. Our simple
experiments show that it depends on the benchmark we consider. The
bottom line is that all our approaches consume less than twice as much
memory as the reference implementation, conservative
version. Moreover, we observe that \textsc{pure-shallow} consumes
10\%-20\% less memory than \textsc{pure-deep}. Concerning the
\textsc{smart} approach, it seems like our Curryfied memoization
scheme creates a lot of very small hash-tables in the OCaml heap,
which is not the optimized use case of OCaml hash-tables. Thus, there
might be room for improvement here.

% \subsection{Maps and sets over hash-consed types}
% We argued in \secref{sec:smart+uid} that it may be necessary to
% implement efficient maps and sets over BDD nodes.
% %
% In this section, we discuss how such maps and sets could be
% implemented in our various approaches.

% First, recall that our \textsc{pure-deep} and \textsc{pure-shallow}
% expose the unique identifiers of the nodes to the user. Since these
% are plain Coq \coqe{positive} numbers, we can use Coq's efficient maps
% and sets over \coqe{positive}.

% Second, the \textsc{smart+uid} approach was designed specifically to
% expose unique identifiers and the suitable order relation in the Coq
% interface as an abstract type, at the expense of being unable to use
% weak hash-tables. This makes it possible to use Coq's efficient finite sets
% and finite maps.
% %
% Alternatively, we argued that one can refrain from exposing the
% comparison function, and expose the signature of maps and sets.

% Finally, let's consider the \textsc{smart} approach. We can actually
% implement a comparison function on BDDs exactly as we implemented the
% operations on BDDs. That is, implement the corresponding ``semantic''
% comparison function on BDTs, and extract it using our memoizing
% fixed-point combinators. This would make it possible to use Coq's
% efficient finite sets and finite maps.

\section{Implementing reduction in the $\lambda$-calculus}\label{sec:lambda}
In the previous sections, we have presented several design patterns
about how to implement hash-consed data structures in Coq. In this
section, we apply two of them to the example of reduction of
$\lambda$-terms (the running example of Conchon and
Filli\^atre~\cite{ConchonFilliatre06wml}).

\subsection{An implementation without hash-consing}
We first describe a Coq implementation, without memoization nor
hash-consing, of $\lambda$-terms with de Bruijn indices inspired from
Huet's \emph{Constructive Computation Theory}~\cite{cct}. We implement four
functions on terms: \coqe{lift}, \coqe{subst}, \coqe{hnf} (that puts a term
in head-normal form) and \coqe{nf} (that puts a term in normal form). This
amounts to roughly 60 lines of Coq code: we present the code for
\coqe{lift} in \figref{fig:lambda-ref-lift} as a reference for future
comparisons.

\begin{figure}[t]
\begin{subfigure}[t]{\textwidth}
\begin{coq}
Inductive term : Type :=
| Var : N -> term
| App : term -> term -> term
| Abs : term -> term.
\end{coq}
\caption{ $\lambda$-terms with de Bruijn indices}\label{fig:term-debruijn}
\end{subfigure}

\begin{subfigure}[t]{\textwidth}
\begin{coq}
Fixpoint lifti (n : N) (t : term) (k : N) :=
  match t with
    | Var i => if N.ltb i k then Var i else Var (i + n)
    | Abs t => Abs ( lifti n t (N.succ k))
    | App t u => App (lifti n t k) (lifti n u k)
  end.
Definition lift n t := lifti n t 0.
\end{coq}
  \caption{A implementation of
    \coqe{lift} without hash-consing nor memoization}\label{fig:lambda-ref-lift}
\end{subfigure}

\begin{subfigure}[t]{\textwidth}
\begin{coq}
Definition lifti : N -> term -> N -> term.
refine (
  memo (fun n =>
    memo_rec (well_founded_ltof _ size) (fun t rec k =>
      memo (fun k =>
        match t with
          | Var i => fun rec => if N.ltb i k then Var i else Var (N.add i n)
          | Abs t => fun rec => Abs (rec t _ (N.succ k) )
          | App t u => fun rec => App (rec t _ k) (rec u _ k)
        end) k rec)));
(* Proof obligations elided *)

Definition lift n t := lifti n t 0.
\end{coq}
\caption{Implementing lift the smart way}\label{fig:lambda-smart-lift}
\end{subfigure}
\caption{Implementation of terms and \coqe{lift}}\label{fig:lambda-defs}
\end{figure}

Then, we reuse Conchon and Filli\^atre's
benchmark~\cite{ConchonFilliatre06wml}: we implement a
$\lambda$-calculus version of quicksort that operates on lists
(encoded as $\lambda$-terms) of Church numerals; and apply this
algorithm to sort the list $L = (0::3::5::2::4::1::nil)$.

Using Coq's virtual machine, sorting this list requires over 250~s on
a recent desktop computer\footnote{Equipped with Intel Xeon cores
  running at 3.60GHz.}: this involves a number of reduction that is
exponential w.r.t. the size of the list and the size of the
numbers.
Extracting the code to OCaml makes it more efficient by a constant
factor. We compile the extracted code with the native OCaml
compiler, and run it to sort the list $L$.
\begin{center}
  \begin{tabular}{|c|c|c|}
    \hline
    time (s) & Number of allocated bytes & Max. size of the major heap
    (words)\\
    \hline
    30 & $47 * 10 ^9$ & $1.1 * 10 ^6$ \\
    \hline
  \end{tabular}
\end{center}
The number of allocated bytes is given as reported by
\ocamle{Gc.allocated_bytes}: it corresponds to what is allocated
during the execution of the sorting algorithm.  The maximum size of
the major heap is given as reported by OCaml \ocamle{Gc.stat}
function. In short, these numbers mean that the program allocates an
awful lot of short-lived values.

\subsection{Using smart constructors}
The \textsc{smart} approach from \secref{sec:smart}
involves a small amount of changes to the code, mainly to use
memoizing constructs and to add \coqe{memoizer} definitions. To get a
rough idea of the changes involved, we present the modified code for
\coqe{lift} in \figref{fig:lambda-smart-lift}: it is merely a matter
of replacing abstractions and fixed-point definitions with their
memoizing counterparts. The overall size of the code is now roughly
over 80 lines of Coq code.
While they do not make it possible to compute efficiently inside Coq,
these changes have dramatic consequences on the execution time of the
extracted code.
Again, we compile the extracted code with the native OCaml compiler,
and run it to sort the list $L$.
\begin{center}
  \begin{tabular}{|c|c|c|}
    \hline
    time (s) & Number of allocated bytes & Max. size of the major heap (words)\\
    \hline
    0.15 &  $0.12 * 10^9$  & $3.1 * 10^6$ \\
    \hline
  \end{tabular}
\end{center}
Here, we have a higher top size of the major heap, which account for
our use of memoization tables. This is to be weighed against the lower
amount of allocated words and the 200$\times$ decrease in running time.
\subsection{Using the shallow-embedding approach}\label{lambda-shallow}
Finally, we use the \textsc{pure-shallow} from
\secref{sec:pure-shallow}. There is now roughly 300 lines of code and
definitions to implement \coqe{lift}, \coqe{subst}, \coqe{hnf} and
\coqe{nf}. In this section, we will make a brief review of this code,
highlighting the differences with respect to what we presented in
\secref{sec:pure-shallow}.

We give the modified definition of terms in \figref{fig:term-shallow}:
each constructor takes an extra \coqe{positive} argument: that is, we
drop the extra indirection we used in the definition of BDD
expressions in \secref{sec:pure-shallow}. (This choice of definition yields
simpler proof arguments because we do not need to perform mutual
inductions on \coqe{expr} and \coqe{hc_expr}.)

We adapt the notion of global state from \secref{sec:pure-shallow}. Again,
the \coqe{hmap} finite map is indexed using a non-recursive comparison
function: it simply compare the head symbols of the terms and, if
equal, compare the tags of the sub-terms. Again, we define one
memoization table per function we wish to memoize.

Then, let us take the example of abstraction to describe term
creation. We define the following smart constructor.
\begin{coq}
Definition mk_Abs t st := mk_term (Abs t 1) st.
\end{coq}
The function \coqe{mk_Abs} is a simple wrapper around \coqe{mk_term}:
it applies the term constructor \coqe{Abs} to the term \coqe{t} and a
dummy unique identifier. Then, \coqe{mk_term} (see right of
\figref{fig:term-mk}) performs a lookup in the hash-consing table. If
there exists a mapping from \coqe{Abs t p} (for some \coqe{p}) to
\coqe{x} in the table, then we return \coqe{x}. Otherwise, the
\coqe{alloc} function replaces the dummy unique identifier with the
next fresh unique identifier available, and it updates the map
\coqe{hmap} with a mapping from the expression \coqe{t} to its tagged
version \coqe{r}.

In effect, this code maintains the invariant that all the terms in the
\coqe{hmap} table are the canonical representative of their
equivalence classes modulo the term comparison used to index
\coqe{hmap}.  That is, we ensure that the terms we build verify the
following property.
\begin{coq}
Definition wf_term st t := get t st.(hmap) = Some t.
\end{coq}

\begin{figure}
  \centering
  \begin{subfigure}{\textwidth}
\begin{multicols}{2}
\begin{coq}
Inductive term : Type :=
| Var : N -> positive -> term
| App : term -> term -> positive -> term
| Abs : term -> positive -> term.


Definition tag t n :=
match t with
   | Var x _ => Var x n
   | App t u _ => App t u n
   | Abs t _ => Abs t n
end.
\end{coq}
\end{multicols}
\caption{Definition of terms, and the associated tagging function}\label{fig:term-shallow}
\end{subfigure}
% ----------------------------------------
\begin{subfigure}{\textwidth}
  \begin{multicols}{2}
\begin{coq}
Definition alloc t (st : state) :=
let r := tag t st.(next) in
(r, mk_state
  {|
    hmap := set t r st.(hmap);
    next := st.(next) + 1
  |} (to_memo st)).

Definition mk_term (t : term) st :=
match get t st.(hmap) with
  | Some x => (x, st)
  | None => alloc t st
end.


$ $
\end{coq}
\end{multicols}
\caption{Updating the global state, and the term constructor}\label{fig:term-mk}
\end{subfigure}

\begin{subfigure}{\textwidth}
\begin{coq}
Let A := (term * N * state).
Let B := (term * state).
Let msr  := fun (n : A) => size (fst (fst n)). (* measure function *)

Definition lifti_rec (n : N) (arg : A) (rec : forall x : A, msr x < msr arg -> B).
refine(
  match arg as arg' return arg = arg' -> B with
  | (t,k,st) => fun Harg =>
    match get (n,ident t,k) (memo_lifti st) with
      | Some t => (t,st)
      | None => let (r,st) :=
                match t  as t' return t = t' -> _ with
                | Var i _ => fun H => if N.ltb i k
                              then mk_Var i st
                              else mk_Var (N.add i n) st
                | Abs t _ => fun H => let (t,st) := rec (t,(N.succ k),st) _  in
                                      mk_Abs t st
                | App t u _ => fun H => let (t,st) := rec (t,k,st) _ in
                                        let (u,st) := rec (u,k,st) _ in
                                        mk_App t u st
                end eq_refl
                in (r,upd_lifti (n,ident t,k) r st)
    end
  end eq_refl).
(* Proof obligations elided *)
Defined.

Definition lifti (n : N) (t : term) (k : N) (st : state) : term * state :=
  Fixm  msr (lifti_rec n) (t,k,st).
\end{coq}
\caption{Definition of \coqe{lifti}}\label{fig:shallow-lifti}
\end{subfigure}
  \caption{$\lambda$-terms revisited, using the shallow-embedding approach}
  \label{fig:lambda-shallow}
\end{figure}

Let us jump to the definition of \coqe{lifti} (see
\figref{fig:shallow-lifti}): it is defined using a well-founded
fixed-point combinator \coqe{Fixm} (elided here) that ensures that the
measure \coqe{msr} is decreasing through recursive calls. The bulk of
this definition is \coqe{lifti_rec}: it is quite similar to the
previous definitions of lift, except that we have to add fancy return
clauses to the match definitions. They are necessary to be able to
prove that the measure actually decreases through recursive
calls. (Also, note that we could have inlined \coqe{lifti_rec} in
\coqe{lifti}, but this two-step definitions makes it for more
palatable proof goals later on).

The upside of these definition is that we are now able to compute the
normal forms of our $\lambda$-terms inside Coq with decent efficiency:
it takes roughly 4~s to sort the list $L$ using \coqe{vm_compute}.
The downside of these definitions is made apparent as soon as we start
to prove that our four functions \coqe{lift}, \coqe{subst}, \coqe{hnf}
and \coqe{nf} enjoy a simple correctness property: that they always return
terms that are well-formed and preserve well-formedness of the global
state. Proving this requires around 600 lines of code, which brings
the total size of this formalization to 900 lines, before we even
started to prove meta-properties like the fact that \coqe{hnf} or
\coqe{nf} implement beta-reduction!

\subsection{Discussion}
In this section, we have demonstrated that our design patterns can be
applied to other settings than BDDs. This exercise helped us to refine
the presentation of smart constructors we gave in \secref{sec:smart},
and gave us the opportunity to present another flavor of the
shallow-embedding technique, that does not rely on the mutually
recursive inductives we used in \secref{sec:pure-shallow}.

In short, this is a nice \emph{exercice de style}. Yet, we believe
that this study is also significant by itself.
Reduction in the $\lambda$-calculus is representative of the kind of
computation that arises in Coq, e.g. in the conversion test. Our case
study can be seen as a step stone for future works that would attempt
to prove the correctness of \emph{efficient} implementations of
symbolic algorithms.
On the one hand, one could use the \textsc{pure-shallow} approach to
implement hash-consing and memoization, yet remain inside the bounds
of what Coq's extraction mechanism can safely handle.
On the other hand, one could bite the bullet, use the \textsc{smart}
approach, and focus on more interesting details than hash-consing and
memoization.

\section{Conclusion}\label{sec:conclusion}
In this paper, we proposed two solutions to implement hash-consing in
programs certified with the Coq system. The first one is to implement
it using Coq data structures; the second is to use the imperative
features provided by OCaml through the tuning of the extraction
mechanism.
The difference in flavor between the mapping of Coq constants to smart
OCaml realizers or the axiomatization of there realizers in Coq is a
matter of taste. In both cases, some meta-theoretical reasoning is
required and requires to ``sweep something under the rug''.

\subsection{Trusting manually-tweaked extracted code}\label{ssec:trust_tweaked_extracted}
Wrapping-up this article, we have to stress the fact that using
hand-written code in the extraction process yields code that is less
trustworthy.
Indeed, the code we use to realize axioms (e.g., inhabitants of the
type \coqe{memoizer}) or to replace constructors and destructors with
the suitable smart ones may introduce bugs.

For instance, we could end up producing code that fails to compile,
because the code snippets used to customize the extraction are used as
textual replacements.
Or, we could end up producing code that is observationally
not-equivalent to the pristine extracted function because of
side-effects.
Or we could simply ruin all our verification effort by introducing the
same kind of bugs that we wanted to catch using formal proofs.

There are two simple ways to restore confidence in this code: code
reviews and tests. Unfortunately, doing more than that is hard.
For instance, we could imagine using the Why3~\cite{Why3} framework to
reason about our code. Unfortunately, Why3 is not suited for
reasoning about higher-order programs, such as our memoizing fixpoint
operators.
More promising is Chargu{\'e}raud's CFML~\cite{chargueraud-11-cfml}
framework, that makes it possible to reason about ML programs with
effects. We wonder to what extent the verification of our combinators
could be conducted in this system.

In any case, we would like to point out that using hand-written code
only \emph{increase} the trusted code base, which contains the whole
extraction mechanism.

\subsection{Trusting external imperative code}

The simple act of calling external libraries that rely on side effects may introduce inconsistencies if these libraries return inconsistent values when called with identical parameters.
This is because, in Coq, a given term has a unique value inside a model (or a trace of execution, in practical terms).
This is the reason why exposing the internals of mutable state, such as unique identifiers, is fraught with possibilities of introducing subtle inconsistencies.

For example, the ``else'' branch of the following Coq program is unreachable, because \coqe{foo x} and \coqe{foo (pred (S x))} are equal, thus the theorem stating that it always returns ``true'':
\begin{coq}
Definition essai (x : nat) : bool :=
  if eq_nat_dec (foo x) (foo (pred (S x))) then true else false.
Theorem essai_true: forall x: nat, (essai x) = true.
\end{coq}
Yet, if foo is extracted to a function such as
\begin{ocaml}
let counter = ref O in fun x -> counter := S (!counter); !counter
\end{ocaml}
then the OCaml code for \coqe{essai O} returns ``false''.

\subsection{Some practical details of implementation}
We took care through this article \emph{not} to use Coq's advanced
support for recursive definitions, i.e., the \textsc{Program} and
\textsc{Function} commands. There are two important reasons for this.
First, as we hinted at in \secref{sec:pure-deep}, the termination
arguments for our BDD operations are out of the scope of what both
commands can automate.
Second, we could have used \textsc{Program} instead of \coqe{refine}
in many of our definitions. Both can be used to build terms with
holes, that are later filled using tactics. However, \textsc{Program}
is based on a powerful
\emph{elaboration}~\cite{DBLP:conf/types/Sozeau06}, meaning that the
final term that is built may be quite different from the one the user
has written.

Generally speaking, the performances of the OCaml version of the
libraries we build using the \textsc{smart} approach are quite
dependent on side-effects.
For instance, keeping memoization tables from one run of a given BDD
operation to the other yields different performance and memory
consumption profiles from what one gets without this conservative
strategy.
The only change between these conservative and non-conservative
strategies is when the memoization tables are allocated. Therefore, we
are quite picky when it comes to program transformations that could
introduce partial applications or eta-expansions. Using \coqe{refine},
we are more confident that the code that we wrote is the one that is
going to be extracted, and we have a better idea of the kind of OCaml
code that is produced. This makes it easier to deal with this kind of
potential performance issues.

\subsection{Related work}
\paragraph{Purely functional implementations of BDDs}
\Textcite{Bradley_Davies_98,DBLP:conf/sfp/ChristiansenH06} implemented \emph{lazy} BDDs in Haskell in a manner similar to our ``pure-shallow'' approach.
In their approaches, not only are BDDs maximally shared, they are also lazily evaluated ``on demand''. We did not investigate this possibility.

\paragraph{Imperative features inside the specification language.}
An obvious solution to implement and reason about imperative
algorithms is to have these imperative features present in the
modeling language of the prover.
Some provers directly target high-level programming languages with
data structures, references and imperatives features: an example is
KeY \cite{KeYBook2007}, which targets a large subset of Java. While
such features are not available in Coq, there are two conceptual
difficulties with this approach that would have made it impractical in
our case-studies.
First, BDD algorithms implemented using hash-consing are functional in
a high-level view: BDD operations are very clearly given functionally,
by induction; but also because hash-consing is suitable only for
immutable structures%
\footnote{Or at least for structures behaving as though they were
  immutable; for instance, we can perform hash-consing on a structure
  if the mutable information in the structure is just used for caching
  and does not affect the hashcode.}.
It therefore seems strange to have to program them in an imperative
language, furthermore one that complicates common functional idioms
(e.g. pattern-matching).  Second, the meta-theory for such languages
is typically huge, with intricate proof rules having to deal with
mutable data, references, late binding etc.  It is not obvious how
much we can trust the proof system with respect to the semantics of
the language.

A second option is to add to an existing functional specification
language certain imperative traits as \emph{monads}, then modify the
extraction function from the specification language to the target
language so as to translate monadic operations into imperative calls,
as has been done for Isabelle/HOL
\cite{Bulwahn:2008:IFP:1459784.1459801}; this approach has been used
to verify a BDD package~\cite{GiorginoSt2012Correctness}. Special
proof idioms have to be used for monadic programs, in addition to the
general difficulty of programming in monadic style
(see~\S\ref{ssec:state_monad}). To the best of our knowledge, this
approach has not been investigated in Coq.

A third option is to integrate into the specification language some
essentially imperative data structures (e.g. mutable arrays, from
which hash tables can be implemented), but present them in a
functional fashion (e.g. an update to a mutable array is treated as
returning a new array, same as the previous except for the updated
location).  The imperative features are then implemented efficiently
for evaluation of expressions inside the prover, and are mapped to
native imperative features of the target language during extraction.
For instance, an experimental version of Coq exists, with native
integers and arrays~\cite{DBLP:conf/itp/ArmandGST10}.  Again, all
difficulties with monads (see~\S\ref{ssec:state_monad}) apply, plus
there is the problem of running a nonstandard version of Coq.

A fourth option is to \emph{deeply embed} a subset of an imperative
language into Coq: the programs of this imperative language are given
a semantics inside Coq, and correctness properties are proved with
respect to this semantics. This idea is discussed by Vafeiadis
\cite{adjustable-references}, but, as he remarks, this largely
precludes the use of regular proof tactics: we have to develop proof
steps specific to the language being embedded, and prove these steps
correct with respect to the semantics.

\paragraph{Verification of BDD algorithms in Coq.}
Verma et al.~\cite{DBLP:conf/asian/VermaGPA00,verma:inria-00072797}
implemented and proved correct in Coq a BDD library featuring
efficient negation and disjunction; other operations like conjunction,
implication and so on are implemented as derived operations.
The BDDs produced are reduced and shared.

As we said, we chose to implement a fresh one because the code
associated to their paper did not age well w.r.t. the evolution of
Coq. Beside the fact that they investigated garbage collection, there
is no conceptual difference between their library and our
\textsc{pure-deep} approach.

\paragraph{Verification of BDD algorithms in other theorem provers}
In Isabelle/HOL, Ortner and Schirmer~\cite{DBLP:conf/tphol/OrtnerS05}
verified the implementation of a normalization algorithm for binary
decision diagrams. That is, their algorithm takes as input a BDD, and
outputs the corresponding ROBDD. Their formalization is built on the
Burstall-Bornat memory model: they build one heap of type $\tt ref
\fmap value$ for each component of a BDD node, with $\tt ref$ the
abstract type of memory addresses. Using a split-heap model makes it
easier to reason about heap-allocated data structures in tools such as
Why3. We believe however, that our formalization is more direct than
theirs, and more suitable for efficient implementations in Coq.

Boyer and Hunt~\cite{DBLP:conf/acl2/BoyerHH06} developed an extension
of ACL2 that uses hash-consing to give canonical representatives to
ACL2 objects. This makes it possible to memoize some ACL2 user defined
functions. As a case study, they implemented a BDD library in
ACL2. Remark that this implementation is based on the fact that ACL2
exposes hash-consing primitives and the associated reasoning
principles to the user. It is unclear to what extent these features
could be added to the Coq proof assistant.

\paragraph{Hash-consing in the execution language}
Jean Goubault-Larrecq's HimML programming language
\cite{jg:HimML94,JG:HimML:dwnld,JG:Implementing94} is an extension of
core Standard~ML with primitive finite set and map data types with a
run-time system designed around the concept of maximal sharing, or
systematic hash-consing.

\subsection{Future work}
We conclude with directions for future work. First, we believe that
the smart constructors approach is generalizable to a huge variety of
inductive types. One can imagine it could be the job of Coq's
extraction mechanism to implement on-demand such smart constructors
and memoizers as it was the case for other imperative
constructs~\cite{DBLP:conf/itp/ArmandGST10}.  Where to stop? Should
the extraction mechanism also provide built-in functional maps and sets
upon hash-consed types?

Second, in \secref{sec:pure-shallow}, we hinted at the fact that using
two mutually inductive data-types for expressions and hash-consed
expressions yields more complicated proofs than the one we had to do
in \secref{lambda-shallow}. However, these two mutually inductive
data-types are at the heart of Conchon and Filli\^atre hash-consing
\emph{library}.
We believe that following their approach closely is a
good starting point for investigating to what extent one could provide
a certified version of this hash-consing library.

Third, we confess that our implementations of BDDs are a bit rough. We
would like to polish the implementations of our \textsc{pure-shallow}
and \textsc{smart} approaches, and release them as user-friendly
libraries. In particular, we need to add some operations like
quantification or functional composition.

%
% In order to use {\OCaml} imperative types in extracted code, we use
% nontrivial semantic properties of the runtime system (sequential
% execution of accesses to the hash table and counter, global storage of
% objects).  These properties are not visible through the pure
% functional semantics of {\OCaml} are thus not through normal {\Coq}
% properties.  We proposed two approaches for using them from {\Coq]:
%   \begin{inparaenum}[1)]
%   \item explicitly, by giving {\Coq} axioms or the behaviour of the
%     {\OCaml} constructs (\secref{sec:axioms})
% \item implicitly, by forcefully mapping inductive {\Coq} types to
%   {\OCaml} types using the extraction facilities
%   (\secref{sec:smart-cons}), with the claims that these types satisfy
%   the metatheoretical axioms of these inductive types.
% \end{inparaenum}
% The same conundrum is faced by those willing to use fast extended
% precision arithmetic: one may either \begin{inparaenum}[1)] \item
%   declare external types for integer, rational etc. and the usual
%   first-order axioms for such types \item force the extraction of
%   {\Coq} arithmetic types to these faat arithmetic
%   types.\end{inparaenum} The difference is philosophical: which
% metatheoretical reasoning is assumed and how to sweep it under the
% rug.
\printbibliography
\end{document}